\documentclass[a4paper]{article} 

\usepackage[a4paper, total={7in, 10in}]{geometry}
\usepackage{graphicx}% Include figure files
\usepackage{dcolumn}% Align table columns on decimal point
\usepackage{bm}% bold math
\usepackage{hhline}
\usepackage[utf8]{inputenc}
\usepackage[T1]{fontenc}
\usepackage{etoolbox}
\usepackage{amsmath}
\usepackage{multirow}
\usepackage{url}
\usepackage{rotating}
\usepackage{hyperref}
\usepackage{orcidlink}

\newcommand{\ten}[1]{\mathrm{e}{#1}}

\title{An Overview of Relativistic Particle Pushers and their Extension to Arbitrary Order Accuracy}
\author{H. Schmitz \orcidlink{0000-0002-1486-064}}
\date{\small{\emph{Central Laser Facility, STFC Rutherford-Appleton Laboratory, Didcot OX11 0QX, United Kingdom} \\
\today}}

\begin{document}
\maketitle

\begin{abstract}
Particle in Cell (PIC) simulations have become a vital tool for the investigation of kinetic processes in plasma physics. Many of the systems investigated with PIC simulations contain particles with relativistic velocities. The correct integration and the knowledge of possible sources of errors in relativistic particle trajectories is of importance to accurately judge the validity of the simulation results. Over the past few decades, various new integration schemes for relativistic particle trajectories in PIC simulations have been proposed. These are aimed at improving numerical accuracy in specific scenarios. This article presents a comprehensive comparison of particle pushers with a focus on explicit schemes. An important class of these schemes is found to be generalisable to arbitrary high order. A comparison of the fourth order variants of these schemes with their second order counterpart is also presented.
\end{abstract}

\section{\label{sec:introduction}Introduction}

Since its inception, the Particle In Cell (PIC) \cite{Dawson:1962,Buneman:1969} method has become an indispensable tool in the investigation of kinetic phenomena in plasma physics. The PIC method consists of two main components, a field solver that calculates the electromagnetic field on a grid, and an integration scheme for the particle orbits, often called the "particle pusher", that calculates the trajectories of the simulation particles. These two components are joined together by suitable interpolation schemes that, on one hand, interpolate the forces of the electromagnetic field from the grid onto the particle positions and, on the other hand, interpolate the particle charges or currents back onto the grid. From the beginning, the PIC technique has been applied to relativistic scenarios \cite{Buneman:1969}. Nowadays relativistic PIC simulations are fundamental in fields ranging from astrophysics \cite{Pohl:2020,Nishikawa:2021} and space and planetary physics \cite{Cerutti:2017,Drake:2006,Hesse:2007} to fusion research and high energy density research \cite{Sentoku:2008,Kemp:2014,Schmit:2014}. Today, many large scale relativistic simulation codes exist, such as Osiris \cite{Fonseca:2002}, EPOCH \cite{Arber:2015}, SMILEI \cite{Derouillat:2018}, WarpX \cite{Vay:2018}, and others.

Relativistic PIC simulations necessitate the need for relativistic particle pushers. The integration method proposed by Boris \cite{Boris:1970} has generally become the dominant scheme for most PIC simulation codes. The Boris scheme is easy to implement, has second order accuracy and very good long time stability, often attributed to the volume preserving nature of the scheme \cite{Qin:2013}. More recently however, it was found that the Boris scheme has some shortcomings in relativistically strong electromagnetic fields. Arefiev et al \cite{Arefiev:2015} found that constants of motion were only poorly conserved for particles accelerated in a high-intensity plane electromagnetic wave. A sub-cycling scheme was proposed based on the instantaneous strength of the magnetic field. Vay \cite{Vay:2008} showed that the calculated trajectory of a relativistic particle propagating in a constant crossed electromagnetic field, such that the total force on the particle vanishes, can deviate from the ideal straight trajectory. He proposed an alternative scheme specifically tailored to produce the correct trajectory in this scenario. As a response to the numerical inaccuracies of the Boris scheme in the relativistic regime, numerous alternative integration schemes have been proposed during the last decade \cite{Higuera:2017,Vay:2008,Petri:2020,Zenitani:2018,Chin:2022,Li:2021,Gordon:2021,Decyk:2023,Zhou:2024,Ripperda:2018}.

Reviews of techniques in PIC simulations have a history almost as old as the PIC simulation scheme itself \cite{Dawson:1983,Birdsall:1985}. Today, many reviews of PIC methods exist focusing on the application in specific fields \cite{Shapoval:2024,Pohl:2020,Nishikawa:2021,Ren:2024,Wu:2019}. There exist also comprehensive reviews and text books on PIC simulations covering topics such as the field solver, the particle interpolation and deposition schemes and the particle pusher \cite{Tskhakaya:2007,Verboncoeur:2005,Godfrey:2014}. Some recent comparisons focusing specifically on different particle integration schemes can be found in Ripperda et al \cite{Ripperda:2018} and Yasir \& Saxena \cite{Yasir:2024}. Ripperda et al \cite{Ripperda:2018} compares the Boris pusher with the integrators proposed by Vay \cite{Vay:2008} and Higuera \& Cary \cite{Higuera:2017} together with their new implicit midpoint method. However, the comparison provides few quantitative results and is limited to a small selection of currently available schemes. Yasir \& Saxena \cite{Yasir:2024} also compare the integrators by Boris, Vay, and Higuera \& Cary but focus mostly on the computational run times of their implementation.

This article aims to provide a comprehensive overview of the numerical accuracy of integration schemes in the relativistic regime. The aim is to classify and compare the various schemes that have been proposed over the last two decades. All schemes considered here are tested in a variety of scenarios and, wherever possible, quantitative measures of the numerical accuracy are given and compared. The hope is to provide guidance to developers and users of simulation codes alike as to which schemes are better suited for a specific scenario. The main focus of the comparison is explicit schemes. These are most often preferred in implementations of the PIC method due to their lower computational cost per time-step. However, one implicit scheme by Ripperda et al \cite{Ripperda:2018} will be included to provide a benchmark.

The paper is organised as follows. Section \ref{sec:integration_schemes} introduces the integration schemes used throughout. A brief numerical outline is presented for some of the schemes. For the remaining schemes, the reader is referred to the respective publications. Section \ref{sec:test_cases} presents the test cases and describes the results. In section \ref{sec:higher_order} a class of "Boris-like" schemes is identified that can be extended to higher order using a method proposed by Yoshida \cite{Yoshida:1993}. Fourth order schemes are implemented and tested against their second order counterparts for a selection of tests. Finally, section \ref{sec:discussion} discusses the results.

\section{\label{sec:integration_schemes}Outline of Integration Schemes}

The relativistic equations of motion for a charged particle in an electromagnetic field are given by
\begin{align}
\frac{d\mathbf{x}}{dt} &= \frac{\mathbf{u}}{\gamma}\label{eq:eqn_motion_x}\\
\frac{d\mathbf{u}}{dt} &= \frac{q}{m}\left(
    \mathbf{E} + \frac{1}{\gamma}\mathbf{u}\times\mathbf{B}
\right)\label{eq:eqn_motion_u}
\end{align}
Here $\mathbf{x}$ is the particle's position and $\mathbf{u} = \gamma \mathbf{v}$ are the spatial components of the four-velocity, with the velocity, $\mathbf{v}$, and the Lorentz factor, $\gamma$. The Lorentz factor can be expressed as a function of the magnitude of the relativistic spatial velocity, $\gamma(u) = (1+u^2/c^2)^{1/2}$. In the following, $\mathbf{u}$ will simply be denoted as the relativistic velocity. The schemes considered in this work can roughly be split into two main types. Type I will denote time centered explicit second order schemes. Type II will denote schemes integrating the trajectory in the particle's proper time. Finally, one implicit scheme by Ripperda et al \cite{Ripperda:2018} will be also considered for comparison.

\subsection{Type I: time centered explicit second order schemes}

An important subset of time-centered second order schemes, which includes the classic Boris scheme \cite{Boris:1970}, the Higuera \& Cary scheme \cite{Higuera:2017}, and Vay's scheme \cite{Vay:2018}, can be expressed as \cite{Higuera:2017},
\begin{align}
    \mathbf{x}^{(n+1/2)} - \mathbf{x}^{(n)} &= \Delta t\frac{\mathbf{u}^{(n)}}{2\gamma(u^{(n)})} \label{eq:scheme_x1}\\
    \mathbf{u}^{(n+1)} - \mathbf{u}^{(n)} &= \Delta t\frac{q}{m}\left(
    \mathbf{E}^{(n+1/2)} + \bar{\mathbf{v}}\times\mathbf{B}^{(n+1/2)} \right) \label{eq:scheme_u}\\
    \mathbf{x}^{(n+1)} - \mathbf{x}^{(n+1/2)} &= \Delta t\frac{\mathbf{u}^{(n+1)}}{2\gamma(u^{(n+1)})}. \label{eq:scheme_x2}
\end{align}
Here, the superscript denotes the time step, $\mathbf{x}^{(n)} = \mathbf{x}(n\Delta t)$, $\mathbf{u}^{(n)} = \mathbf{u}(n\Delta t)$ and $\mathbf{E}^{(n)}=\mathbf{E}(\mathbf{x}^{(n)}, n\Delta t)$, $\mathbf{B}^{(n)}=\mathbf{B}(\mathbf{x}^{(n)}, n\Delta t)$ and $\Delta t$ is the step size of the scheme. In the above expression, $\bar{\mathbf{v}}$ is a suitably averaged velocity. As has been pointed out in \cite{Higuera:2017}, this allows for different schemes depending on how $\bar{\mathbf{v}}$ is calculated. The classic Boris scheme uses \cite{Boris:1970}
\begin{displaymath}
\bar{\mathbf{v}}_{\mathrm{B}} = \frac{1}{2}\left(
    \mathbf{v}(\mathbf{u}^{(n)} + \pmb{\epsilon}^{(n+1/2)}) + \mathbf{v}(\mathbf{u}^{(n+1)} - \pmb{\epsilon}^{(n+1/2)})
\right),
\end{displaymath}
with
\begin{displaymath}
\pmb{\epsilon}^{(n+1/2)} = \frac{q \Delta t}{2 m}\mathbf{E}^{(n+1/2)}.
\end{displaymath}
Here $\mathbf{v}(\mathbf{u})$ denotes the velocity $\mathbf{v}$ expressed as a function of the relativistic velocity $\mathbf{u}$,
\begin{displaymath}
    \mathbf{v}(\mathbf{u}) = \frac{\mathbf{u}}{\sqrt{1 + \left(u/c\right)^2}}.
\end{displaymath}
This choice separates the velocity updates from the electric and magnetic field, resulting in the well known split update steps. This split has the consequence that, in the case that electric and magnetic forces cancel, the scheme is not guaranteed to reproduce a force-free trajectory. For this reason Vay \cite{Vay:2008} proposes the choice
\begin{displaymath}
\bar{\mathbf{v}}_{\mathrm{V}} = \frac{1}{2}\left(
    \mathbf{v}(\mathbf{u}^{(n)}) + \mathbf{v}(\mathbf{u}^{(n+1)})
\right).
\end{displaymath}
This, however, results in a scheme that is not volume preserving in $(\mathbf{x}, \mathbf{u})$ phase-space. Therefore, Higuera \& Cary \cite{Higuera:2017} use
\begin{displaymath}
\bar{\mathbf{v}}_{\mathrm{HC}} = \mathbf{v}\left(\frac{1}{2}
    (\mathbf{u}^{(n)} + \mathbf{u}^{(n+1)})
\right).
\end{displaymath}
All three schemes, Boris \cite{Boris:1970}, Vay \cite{Vay:2008}, and Higuera \& Cary \cite{Higuera:2017} can be written as explicit updates 
\begin{displaymath}
(\mathbf{x}^{(n+1)}, \mathbf{u}^{(n+1)}) = \phi_{\Delta t}(\mathbf{x}^{(n)}, \mathbf{u}^{(n)}),     
\end{displaymath}
where $\phi_{\Delta t}$ can be seen as a function that maps the six-dimensional phase space onto itself. Because the updates in equations (\ref{eq:scheme_x1} -- \ref{eq:scheme_x2}) can be performed sequentially, this map can be split into three parts,
\begin{equation}
    \phi_{\Delta t} = T^x_{\Delta t / 2}  \phi^u_{\Delta t}  T^x_{\Delta t / 2},\label{eq:op_scheme}
\end{equation}
where $T^x_{\Delta t / 2}$ is a translation in the coordinates depending only on $\mathbf{u}$ and $\phi^u_{\Delta t}$ is the explicit velocity update derived from equation (\ref{eq:scheme_u}) whose exact form depends on the chosen scheme. 

For Boris and Higuera \& Cary, this velocity update can be split up further. In both cases, the update in equation (\ref{eq:scheme_u}) can be written in three steps
\begin{align}
    \mathbf{u}^- &= \mathbf{u}^{(n)} + \pmb{\epsilon}^{(n+1/2)}\label{eq:scheme_e1}\\
    \mathbf{u}^+ - \mathbf{u}^- &= \frac{1}{\gamma_s}\left(
        \mathbf{u}^+ + \mathbf{u}^-
    \right)\times \pmb{\beta}^{(n+1/2)}\label{eq:scheme_b}\\
    \mathbf{u}^{(n+1)} &= \mathbf{u}^+ + \pmb{\epsilon}^{(n+1/2)}, \label{eq:scheme_e2}
\end{align}
where 
\begin{equation}
\pmb{\beta}^{(n+1/2)} = \frac{q \Delta t}{2 m}\mathbf{B}^{(n+1/2)},\label{eq:def_beta}
\end{equation}
and $\gamma_s$ is a Lorentz factor that depends on the scheme $s$. For Boris it is given by
\begin{displaymath}
    \gamma_s = \gamma_{\mathrm{B}} = \gamma_{-} = \sqrt{1 + \left(\frac{u^-}{c}\right)^2},
\end{displaymath}
while Higuera \& Cary give
\begin{displaymath}
    \gamma_s = \gamma_{\mathrm{HC}} = \sqrt{
    \frac{1}{2}\left(\gamma_{-}^{2}-\beta^{2}+\sqrt{\left(\gamma_{-}^{2}-\beta^{2}\right)^{2}+4\left(\beta^{2}+\left|\pmb{\beta} \cdot \mathbf{u}_{-}\right|^{2}\right)}\right)
    }
\end{displaymath}
Note that, in the limit of small time steps, $\gamma_{\mathrm{HC}}$ converges to $\gamma_{\mathrm{B}}$. Equation (\ref{eq:scheme_b}) can be shown to be a pure rotation of $\mathbf{u}$ around the magnetic field direction by an angle $\theta_s$ given by 
\begin{equation}
    \tan\left(\frac{\theta_s}{2}\right) = \frac{\beta}{\gamma_s}. \label{eq:rot_angle_boris}
\end{equation}
Here $\beta = |\pmb{\beta}|$ is given by equation (\ref{eq:def_beta}). Thus, we can split $\phi^u_{\Delta t}$ up further,
\begin{equation}
    \phi^u_{\Delta t} = T^E_{\Delta t/2}\left(\pmb{\epsilon}\right) R^B_{\Delta t}\left(\theta_s\right) T^E_{\Delta t/2}\left(\pmb{\epsilon}\right),\label{eq:op_velocity}
\end{equation}
where $T^E_{\Delta t/2}\left(\pmb{\epsilon}\right)$ is a translation in velocity due to the electric field and $R^B_{\Delta t}\left(\theta_s\right)$ is a rotation in velocity due to the magnetic field. Thus, both Boris' and Higuera \& Cary's schemes can be written as a sequence of translations and rotations in $(\mathbf{x}, \mathbf{u})$ phase-space. Each of these transformations is trivially volume-preserving and, as a consequence, both schemes preserve phase-space volume overall. However, as Qin et al \cite{Qin:2013} have pointed out, Boris' scheme is not symplectic which is a stronger condition that is often linked to long-term stability of a numerical scheme. A similar calculation as carried out there reveals that Higuera \& Cary's scheme also is not symplectic.

The rotation angle $\theta_s$ given in equation (\ref{eq:rot_angle_boris}) suggests that a more exact scheme might be obtained if the accuracy of the scheme is improved by replacing this angle with the exact value of the cyclotron motion for a particle gyrating in a constant magnetic field,
\begin{equation}
    \theta_{\mathrm{C}}= \frac{2\beta}{\gamma_B}. \label{eq:rot_angle_exact}
\end{equation}
As Zenitani and Umeda \cite{Zenitani:2018} have pointed out, the scheme defined by equations (\ref{eq:op_scheme}) and (\ref{eq:op_velocity}), where $\theta_s$ is given by $\theta_{\mathrm{C}}$ in equation (\ref{eq:rot_angle_exact}) was already proposed in \cite{Boris:1970}. 
It can be described as gyrophase correction to the well known Boris-scheme. The version used here follows the implementation in \cite{Zenitani:2018} that uses Rodrigues' rotation formula \cite{Goldstein:1980} and will be denoted by GYR throughout the remainder of this article. It is not generally considered a viable scheme in the context of PIC simulations due to the need to evaluate trigonometric functions at every time step for every particle which can harm the performance of a simulation code. This scheme is included for comparison and to allow the analysis of the origin of approximation errors in the tests. By construction, the phase error of GYR in a constant magnetic field is exactly zero. Thus, errors attributed to phase errors can be analysed. Umeda \cite{Umeda:2018} proposed a Boris-like pusher that uses a three step process to perform the velocity rotation. This provides an improvement in the accuracy of the rotation angle at moderate computational cost. However, computational efficiency is not the primary concern of this article, 
and modern computer architectures may allow for more computationally intensive algorithms without increasing overall wall-time \cite{Decyk:2014}. Since GYR allows the comparison with a scheme using the exact rotation angle, Umeda's three step integrator is not included in the current work.

Chin \& Cator \cite{Chin:2022} describe another algorithm, denoted by CC, in which the angle of rotation is given by
\begin{equation}
    \sin\left(\frac{\theta_{\mathrm{CC}}}{2}\right) = \frac{\beta}{\gamma_s}. \label{eq:rot_angle_cc}
\end{equation}
This angle is chosen such that the radius of gyration of the numerical trajectory is exactly the Larmor radius given by 
\begin{displaymath}
    R_g = \frac{q u B}{\gamma m}.
\end{displaymath}
Note that Chin \& Cator \cite{Chin:2022} only describe the integrators in the non-relativistic limit, but the generalisation to relativistic velocities is trivial. The CC pusher is included in the comparison because, in contrast to the traditional Boris pusher and most of its variations, it overestimates the rotation rotation angle rather than underestimating it. 

As stated above, the methods defined by equations (\ref{eq:op_scheme}) and (\ref{eq:op_velocity}) are volume-preserving and time-reversible. The second property renders all of these schemes second order accurate. They are, however, not symplectic. Patacchini and Hutchinson \cite{Patacchini:2009} present a \emph{cyclotronic integrator} that is symplectic and time-reversible in the case of a static, homogeneous magnetic field. Their integrator is excluded from any comparisons in this article because of this limitation and the fact that many of the tests contain time-dependent or non-homogeneous fields.

\subsection{Type II: integration of particle orbits in proper time}

A different class of integration methods that have gained attention over recent years is based on the fact that the relativistic equations of motion in a constant but arbitrary electromagnetic field can be solved exactly in the particle's proper time $\tau$, i.e. the time measured in the instantaneous rest frame of the particle and related to the simulation time by
\begin{displaymath}
    \frac{d\tau}{dt} = \frac{1}{\gamma}.
\end{displaymath}
The difficulty then arises in determining the proper time step size $\Delta \tau$ for a given $\Delta t$ in the simulation frame of reference. This requires solving an implicit equation.
P\'etri \cite{Petri:2020} presents a method that solves the equations of motion exactly using the proper time and then uses Newton-Raphson's method to solve the implicit time equation up to some predetermined precision. His solution involves transforming into a frame in which the electric and magnetic fields are parallel.
The method of P\'etri et al can be considered exact up to machine precision in the case that the fields are constant.  Li et al \cite{Li:2021} provide a simpler formulation of the exact integrator using the covariant form of the equations of motion. They define
\begin{equation}
\kappa=\frac{1}{\sqrt{2}} \sqrt{\mathcal{I}_1+\sqrt{\mathcal{I}_1^2+4 \mathcal{I}_2^2}}, \quad \sigma=\frac{1}{\sqrt{2}} \sqrt{-\mathcal{I}_1+\sqrt{\mathcal{I}_1^2+4 \mathcal{I}_2^2}},
\end{equation}
where 
\begin{equation}
    \mathcal{I}_1=|\mathbf{E}|^2-|\mathbf{B}|^2 \quad \text{and} \quad \mathcal{I}_2=\mathbf{E} \cdot \mathbf{B}.
\end{equation}
Note here the symbol $\sigma$ is used where \cite{Li:2021} uses $\omega$. This change in notation is chosen to avoid confusion with the angular frequency used later in this article. The motion is decomposed into two independent parts by using the projection operators
\begin{equation}
P_\kappa=\frac{\sigma^2 I+F^2}{\kappa^2+\sigma^2}, \quad \text{and} \quad P_\sigma=\frac{\kappa^2 I-F^2}{\kappa^2+\sigma^2},
\end{equation}
where $F$ is the field tensor and $I$ is the identity matrix. Projection of the four-position $x$ gives $x_\kappa = P_\kappa x$ and $x_\sigma = P_\sigma x$, and likewise for the four-velocity $u$. The exact equations of motion can then be written as \cite{Li:2021}
\begin{equation}
\begin{aligned}
x^{(n+1)} &= x^{(n)} + x_\kappa(\Delta \tau) + x_\sigma(\Delta \tau)\\
x_\kappa(\Delta \tau) & =\left[u_{\kappa}^{(n)} \operatorname{sinc}(\mathrm{i} \kappa \Delta  \tau)+\frac{1}{2} F u_{\kappa}^{(n)} \operatorname{sinc}^2\left(\frac{\mathrm{i} \kappa \Delta \tau}{2}\right) \Delta \tau\right] \Delta \tau, \\
x_\sigma(\Delta \tau) & =\left[u_{\sigma}^{(n)} \operatorname{sinc}(\sigma \Delta \tau)+\frac{1}{2} F u_{\sigma}^{(n)} \operatorname{sinc}^2\left(\frac{\sigma \Delta \tau}{2}\right) \Delta \tau\right] \Delta \tau,
\end{aligned}\label{eq_exact_x}
\end{equation}
for the four-position $x$ and 
\begin{equation}
\begin{aligned}
u^{(n+1)} &= u_\kappa(\Delta \tau) + u_\sigma(\Delta \tau)\\
u_\kappa(\Delta \tau) &= u_{\kappa}^{(n)} \cosh (\kappa \Delta \tau)+F u_{\kappa}^{(n)} \operatorname{sinc}(\mathrm{i} \kappa \Delta \tau) \Delta \tau, \\
u_\sigma(\Delta \tau) &= u_{\sigma}^{(n)} \cos (\sigma\Delta  \tau)+F u_{\sigma}^{(n)} \operatorname{sinc}(\sigma \Delta \tau) \Delta \tau
\end{aligned}\label{eq_exact_u}
\end{equation}
for the four-velocity $u$. Here $\tau$ is the proper time, $x^{(n)}$ is the four-position and $u_{\kappa}^{(n)}$ and $u_{\sigma}^{(n)}$ are the projections of the four-velocity at time $t^{(n)}$. The relation between simulation time and proper time can be obtained from the zero-component of the four-position,
\begin{equation}
\begin{aligned}
c \Delta t = &\biggl( (u_{\kappa}^{(n)})^0 \operatorname{sinc}(\mathrm{i} \kappa \Delta \tau)+(u_{\sigma}^{(n)})^0 \operatorname{sinc}(\sigma \Delta \tau)\\
& +\frac{\mathbf{u}^{(n)} \cdot \mathbf{E}}{2}\left[\operatorname{sinc}^2\left(\frac{\mathrm{i} \kappa \Delta \tau}{2}\right)+\operatorname{sinc}^2\left(\frac{\sigma \Delta \tau}{2}\right)\right] \Delta \tau\biggr) \Delta \tau.
\end{aligned}\label{eq_proper_time}
\end{equation}
For a given simulation time step $\Delta t$, equation (\ref{eq_proper_time}) defines an implicit relation for the proper time step $\Delta \tau$. Li et al \cite{Li:2021} propose to solve this numerically using a root-finding algorithm such as Newton-Raphson. An alternative scheme is given, where equation (\ref{eq_exact_u}) is used as a leapfrog step together with a position update that assumes a constant velocity. Gordon \& Hafizi \cite{Gordon:2021} present a somewhat more elegant derivation for the velocity update using Pauli matrices and making use of the structure of the Lorentz group $\mathsf{SO}(3,1)$. 
Their update of the four-velocity is exact for constant fields but uses a leap-frog method to update the position. The main focus of the Gordon \& Hafizi pusher is the solution of the trajectory in proper time. 
They suggest using a second-order approximation of equation (\ref{eq_proper_time}) for the proper time step $\Delta \tau$. In contrast, Zhou and Zhang \cite{Zhou:2024} solve the equations of motion (\ref{eq_exact_x}) and (\ref{eq_exact_u}) using a second-order central difference scheme that is solved in a manner similar to the Boris scheme. Thus, while integrating the orbit in proper time, their scheme is at most second order. The interpolation of the time step is then carried out using Newton-Raphson's method. An optimal region is given such that Newton-Raphson's method is guaranteed to converge. 
Another integration scheme using the particle's proper time $\tau$ to calculate an exact solution, presented by Decyk et al \cite{Decyk:2023} is only mentioned here, as it is generally outperformed by the simpler and earlier integrator presented in \cite{Li:2021}. More recently, a Lorentz invariant, volume and energy preserving scheme, VELPA was proposed \cite{RuiliZhang:2024}. However, this scheme is not suitable as a general purpose scheme since it was specifically developed for the strong magnetic field regime $c |\mathbf{B}| \gg |\mathbf{E}|$. In addition, the solution is expressed in terms of proper time only and does not contain a prescription on how to map a given simulation time step $\Delta t$ to proper time $\Delta \tau$.  

% The pusher presented by P\'etri \cite{Petri:2020} and Li et al \cite{Li:2021} is exact in the case of constant fields. All other schemes presented in this section second order for this case. For non-constant fields, the order of all of these schemes is reduced as demonstrated later in this article. For the "exact" pusher, the fields have to be evaluated at time $t^(n)$ and position $x^(n)$ leading to a first-order scheme. All other schemes make use of leaf-frogging but they still require the inversion of equation (\ref{eq_proper_time}), or an approximation thereof, which contains the product $\mathbf{u}^{(n)} \cdot \mathbf{E}$. Without giving proof it is conjectured that this cannot be done in such a way as to produce a time-reversible scheme.

\subsection{Implicit midpoint method}

Although, the integrators chosen for the comparisons in this article are mainly explicit, one implicit method will be included in the comparison. 
The choice of method to include is somewhat arbitrary as there have been numerous algorithms proposed over the past years \cite{Petri:2017,He:2015,Lapenta:2011a,Ripperda:2018}. The Implicit Midpoint Method presented by Ripperda et al \cite{Ripperda:2018} stands out as one of the simplest volume preserving schemes. Their method uses a time centered expression,
\begin{align}
    \mathbf{x}^{(n+1)} - \mathbf{x}^{(n)} &= \Delta t\bar{\mathbf{v}}_{\mathrm{IMP}}\\
    \mathbf{u}^{(n+1)} - \mathbf{u}^{(n)} &= \Delta t\frac{q}{m}\left(
        \mathbf{E}^{(n+1/2)} + \bar{\mathbf{v}}_{\mathrm{IMP}}\times\mathbf{B}^{(n+1/2)} 
    \right)\\
    \bar{\mathbf{v}}_{\mathrm{IMP}} &= \frac{\mathbf{u}^{(n+1)} + \mathbf{u}^{(n)}}{\gamma^{(n+1)} + \gamma^{(n)}}.
\end{align}
These equations are solved for $\mathbf{u}^{(n+1)}$ by starting from an initial guess and iterating using the Jacobian of a residual error. The Implicit Midpoint Method has been shown to be second order accurate \cite{Ripperda:2018}.

\subsection{Properties of the integration schemes}

\begin{table}
    \centering
    \begin{tabular}{|c|c|c|}\hline
         {\bfseries Type} & {\bfseries Scheme} & {\bfseries Description} \\ \hline
         \multirow{5}{*}{I}
         & Boris & Classic scheme by Boris \cite{Boris:1970}\\
         & Vay & Vay \cite{Vay:2008} \\
         & HC & Higuera \& Cary \cite{Higuera:2017}\\
         & GYR & Boris with exact rotation \cite{Boris:1970,Zenitani:2018} \\
         & CC & Chin \& Cator \cite{Chin:2022}\\ \hline
         \multirow{5}{*}{II}
         & PL & Analytical scheme by P\'etri \cite{Petri:2020} and Li et al \cite{Li:2021} \\
         & LiLF & Leapfrog scheme by Li et al \cite{Li:2021} \\
         & GH & Gordon \& Hafizi \cite{Gordon:2021}, exact propagator \\
         & GH2 & Gordon \& Hafizi \cite{Gordon:2021}, $2^\mathrm{nd}$ order propagator \\
         & ZZ & Zhou \& Zhang \cite{Zhou:2024} \\ \hline
         & IMP & Implicit Midpoint Method, Ripperda et al \cite{Ripperda:2018} \\ \hline
    \end{tabular}
    \caption{List of the numerical schemes used throughout this article.}
    \label{tab:schemes}
\end{table}

Table \ref{tab:schemes} lists the integration schemes used throughout this article together with their references. In this section, some of the known properties of the schemes are summarised.

\emph{Order of convergence:} The order of convergence denotes the asymptotic behaviour of the error in the limit $\Delta t \to 0$. All type I schemes, Boris, Vay, HC, GYR, and CC, as well as the implicit scheme IMP, show second order convergence independent of the field configuration. For constant, homogeneous fields, PL is exact and the order of convergence can be denoted as infinite. All other type II schemes are second order in this case. As will be seen later in this article, the convergence of type II schemes can reduce to first order when the electromagnetic fields vary in space or time. For the non-staggered scheme PL, this can be understood from the fact that the field tensor $F$ has to be evaluated at time $t^{(n)}$ in equations (\ref{eq_exact_x}) and (\ref{eq_exact_u}). This results in a non time-reversible scheme. The type II schemes LiLF, GH, GH2, and ZZ stagger the position and velocity update in time. But equation (\ref{eq_proper_time}) to determine the proper time contains the product $\mathbf{u}\cdot\mathbf{E}(\mathbf{x}, t)$. This term can not be evaluated in a time-centered manner leading to a decrease of the order of convergence for these schemes.

\emph{Symplecticity and volume preservation:} PL is exact for constant, homogeneous fields, and therefore, by definition, symplectic and volume preserving in this case. None of the other schemes claim to be symplectic. For some schemes is has been shown that they are not symplectic, e.g. \cite{Qin:2013} for the case of the Boris scheme. The type I schemes Boris, HC, GYR, and CC can be decomposed according to equations (\ref{eq:op_scheme}) and (\ref{eq:op_velocity}). Each of the individual transformations is trivially volume preserving in $(\mathbf{x}, \mathbf{u})$-space and, subsequently, so is their composition. The Vay scheme is known not to be volume preserving \cite{Ripperda:2018}. The implicit scheme IMP has been constructed to be volume preserving \cite{Ripperda:2018}. No statement is made about the volume preservation properties of the type II schemes other than PL in the constant, homogeneous field case.

\emph{Energy conservation:} In the case of pure magnetic fields $\mathbf{E}=0$, the schemes presented in this article are expected to be energy conserving. However, all schemes presented here sample the electric and magnetic field at discrete locations in space and time. They do not incorporate any information of the potentials into their formalisms which means that they cannot conserve energy in the general case.

\section{\label{sec:test_cases}Test cases}

A variety of tests have been performed, comparing all integration schemes against each other. The test cases described in the following sections are as follows. (A) Gyromotion with $\mathbf{E}=0$ and constant $\mathbf{B}$ (section \ref{sec:gyromotion}), (B) lorentz-boosted particle at rest, $\mathbf{E} = -\mathbf{v}_0\times\mathbf{B}$ (section \ref{sec:lorentz_boost}),
(C) gyration in a Lorentz-boosted frame (section \ref{sec:exb_drift}),
(D) particle oscillating in spatially varying $\mathbf{E} \| \mathbf{B}$ proposed in \cite{Higuera:2017} (section \ref{sec:hg_test}),
(E) magnetic bottle configuration (section \ref{sec:magnetic_bottle}),
(F) acceleration of a particle in a plane wave (section \ref{sec:rel_laser}), and
(G) oscillating electric field $\mathbf{E} \| \mathbf{B}$ (section \ref{sec:oscillating_ez}). For each case, convergence tests in $\Delta t$ were performed for all schemes. Unless otherwise stated, the expected second order convergence was observed. Table \ref{tab:error_measures} summarises the accuracy of the schemes by giving the values of numerical errors of each scheme. All schemes were implemented in Python using double precision floating point numbers. The schemes were not optimised for speed, but clarity of the implementation was preferred instead. For this reason, the current investigation does not allow for a rigorous comparison of the computational cost of the schemes. Vector and matrix operations were implemented using NumPy and root finding operations used the \texttt{brentq} function provided by the SciPy library with \texttt{xtol=1e-30} and \texttt{rtol=1e-15}. The IMP scheme was implemented with a relative tolerance of $1e-13$ and a maximum of 25 iterations. Errors with a magnitude below $10^{-13}$ are considered numerical rounding errors and reported as zero.

%\clearpage
\begin{sidewaystable}
    \centering
    \resizebox{0.75\textwidth}{!}{
    \begin{tabular}{|>{\bfseries}c|l|>{\footnotesize}c|>{\footnotesize}c|>{\footnotesize}c|>{\footnotesize}c|>{\footnotesize}c|>{\footnotesize}c|>{\footnotesize}c|>{\footnotesize}c|>{\footnotesize}c|>{\footnotesize}c|>{\footnotesize}c|}\hline
&  &  \textbf{Boris }& \textbf{Vay}    & \textbf{HC}     & \textbf{GYR}    & \textbf{CC}     & \textbf{PL}     &  \textbf{LiLF}  & \textbf{GH}     & \textbf{GH2}    & \textbf{ZZ}     & \textbf{IMP}    \\ \hline
\multirow{4}{*}{A}
& $\mathcal{E}_\phi^{\mathrm{G}} (a)$
    & 0.1953          & 0.1953          & 0.1948          & \textbf{0}      & -0.1082         & \textbf{0}      & \textbf{0}      & \textbf{0}      & $5.093\ten{-2}$ & 0.1953          & 0.1953          \\ \cline{2-13}
& $\mathcal{E}_u^{\mathrm{G}} (a)$
    & \textbf{0}      & \textbf{0}      & \textbf{0}      & \textbf{0}      & \textbf{0}      & \textbf{0}      & \textbf{0}      & \textbf{0}      & \textbf{0}      & \textbf{0}      & \textbf{0}      \\ \cline{2-13}
& $\mathcal{E}_\phi^{\mathrm{G}} (b)$
    & 0.1953          & 0.1953          & -0.1046         & \textbf{0}      & -0.1082         & \textbf{0}      & \textbf{0}      & \textbf{0}      & $5.093\ten{-2}$ & 0.1953          & 0.1953          \\ \cline{2-13}
& $\mathcal{E}_u^{\mathrm{G}} (b)$
    & \textbf{0}      & \textbf{0}      & \textbf{0}      & \textbf{0}      & \textbf{0}      & \textbf{0}      & \textbf{0}      & \textbf{0}      & \textbf{0}      & \textbf{0}      & \textbf{0}      \\ 
\hhline{|=|=|=|=|=|=|=|=|=|=|=|=|=|}
\multirow{8}{*}{B} 
& $\mathcal{E}_\phi^{\mathrm{L}}   (a)$
    & $9.852\ten{-5}$ & \textbf{0}      & \textbf{0}      & $3.337\ten{-2}$ & $5.329\ten{-2}$ & \textbf{0}      & \textbf{0}      & \textbf{0}      & \textbf{0}      & \textbf{0}      & \textbf{0}      \\ \cline{2-13}
& $\mathcal{E}_\gamma^{\mathrm{L}} (a)$
    & $3.939\ten{-4}$ & \textbf{0}      & \textbf{0}      & 0.1279          & 0.1922          & \textbf{0}      & \textbf{0}      & \textbf{0}      & \textbf{0}      & \textbf{0}      & \textbf{0}      \\ \cline{2-13}
& $\mathcal{E}_\phi^{\mathrm{L}}   (b)$
    & 0.1420          & \textbf{0}      & \textbf{0}      & $7.556\ten{-2}$ & $4.955\ten{-3}$ & \textbf{1.096e-12} & \textbf{1.096e-12} & \textbf{0}      & \textbf{0}      & \textbf{0}      & \textbf{0}      \\ \cline{2-13}
& $\mathcal{E}_\gamma^{\mathrm{L}} (b)$
    & 10.2781         & \textbf{0}      & \textbf{0}      & 3.4456          & 0.1089          & \textbf{1.211e-11} & \textbf{1.211e-11} & \textbf{0}      & \textbf{0}      & \textbf{0}      & \textbf{0}      \\ \cline{2-13}
& $\mathcal{E}_\phi^{\mathrm{L}}   (c)$
    & 0.1501          & \textbf{0}      & \textbf{0}      & $8.800\ten{-2}$ & $4.951\ten{-4}$ & \textbf{1.419e-11} & \textbf{1.419e-11} & \textbf{0}      & \textbf{0}      & \textbf{0}      & \textbf{0}      \\ \cline{2-13}
& $\mathcal{E}_\gamma^{\mathrm{L}} (c)$
    & 945.6164        & \textbf{1.552e-12} & \textbf{2.919e-12} & 321.5242        & 0.1000          & $5.857\ten{-8}$ & $5.857\ten{-8}$ & \textbf{2.952e-12} & \textbf{6.833e-13} & \textbf{0}      & \textbf{0}     \\ \cline{2-13}
& $\mathcal{E}_\phi^{\mathrm{L}}   (d)$
    & 0.1501          & \textbf{0}      & \textbf{0}      & $8.815\ten{-2}$ & $4.951\ten{-5}$ & $8.614\ten{-9}$ & $8.614\ten{-9}$ & \textbf{0}      & \textbf{0}      & \textbf{0}      & \textbf{0}      \\ \cline{2-13}
& $\mathcal{E}_\phi^{\mathrm{L}}   (e)$
    & 0.1501          & \textbf{1.313e-12} & \textbf{1.055e-12} & $8.816\ten{-2}$ & $4.951\ten{-6}$ & \textbf{1.917e-12} & \textbf{1.917e-12} & \textbf{8.769e-13} & \textbf{1.115e-12} & \textbf{0}      & \textbf{1.375e-12} \\
    \hhline{|=|=|=|=|=|=|=|=|=|=|=|=|=|}
\multirow{4}{*}{C}
 & $\mathcal{E}_t^{\mathrm{ExB}} (a)$        
    & $2.329\ten{-3}$ & $6.216\ten{-4}$ & $-1.091\ten{-5}$ & $4.504\ten{-6}$ & $2.624\ten{-6}$ & \textbf{9.391e-12} & \textbf{9.391e-12} & $4.432\ten{-4}$ & $4.491\ten{-4}$ & $4.568\ten{-4}$ & $4.568\ten{-4}$ \\ \cline{2-13}
 & $\mathcal{E}_x^{\mathrm{ExB}} (a)$        
    & $-3.506\ten{-6}$ & \textbf{-3.001e-13} & $-2.891\ten{-5}$ & $-1.550\ten{-5}$ & $-2.115\ten{-5}$ & \textbf{-2.905e-13} & $2.031\ten{-5}$ & $2.519\ten{-5}$ & $2.470\ten{-5}$ & \textbf{-3.044e-13} & \textbf{-3.036e-13} \\ \cline{2-13}
 & $\mathcal{E}_t^{\mathrm{ExB}} (b)$        
    & $6.209\ten{-2}$ & 47.5591         & 44.8069         & 4.7202          & 17.2083         & \textbf{9.797e-10} & \textbf{9.797e-10} & -1.1922         & -2.2432         & $4.724\ten{-4}$ & $4.794\ten{-4}$ \\ \cline{2-13}
 & $\mathcal{E}_x^{\mathrm{ExB}} (b)$
    & $1.186\ten{-4}$ & $-1.019\ten{-11}$ & $-5.421\ten{-6}$ & $9.324\ten{-6}$ & $-2.612\ten{-5}$ & \textbf{1.250e-12} & $-7.245\ten{-5}$ & $6.179\ten{-2}$ & $9.355\ten{-2}$ & \textbf{1.281e-12} & \textbf{1.280e-12} \\ 
    \hhline{|=|=|=|=|=|=|=|=|=|=|=|=|=|}
\multirow{3}{*}{D}
 & $\mathcal{E}^{\mathrm{HC}}_{H}$
     & $5.050\ten{-4}$ & $8.012\ten{-4}$ & $5.050\ten{-4}$ & $5.050\ten{-4}$ & $5.050\ten{-4}$ & $6.288\ten{-2}$ & $1.381\ten{-2}$ & $3.939\ten{-2}$ & $3.935\ten{-2}$ & $1.950\ten{-4}$ & \textbf{0}      \\ \cline{2-13}
 & $\mathcal{E}^{\mathrm{HC}}_{I}$
     & \textbf{0}      & $5.531\ten{-3}$ & \textbf{0}      & \textbf{0}      & \textbf{0}      & \textbf{0}      & \textbf{0}      & \textbf{0}      & \textbf{0}      & \textbf{0}      & \textbf{0}      \\ \cline{2-13}
 & $\mathcal{E}^{\mathrm{HC}}_{p}$
     & $1.742\ten{-3}$ & $4.439\ten{-4}$ & $2.435\ten{-3}$ & $2.586\ten{-3}$ & $3.008\ten{-3}$ & 0.2023          & $5.601\ten{-2}$ & 0.1431          & 0.1431          & $4.317\ten{-4}$ & \textbf{0}      \\ \hhline{|=|=|=|=|=|=|=|=|=|=|=|=|=|}
\multirow{2}{*}{E}
 & $\mathcal{E}^{\text{Mirror}}_\gamma$       
     & \textbf{0}      & $1.390\ten{-3}$ & \textbf{0}      & \textbf{0}      & \textbf{0}      & \textbf{0}      & \textbf{0}      & \textbf{0}      & \textbf{0}      & \textbf{0}      & \textbf{0}      \\ \cline{2-13}
 & $\mathcal{E}^{\text{Mirror}}_\mu$      
     & \textbf{8.395e-8} & $3.228\ten{-2}$ & $3.708\ten{-7}$ & $5.123\ten{-7}$ & $7.690\ten{-7}$ & 0.1308          & $6.053\ten{-7}$ & $6.053\ten{-7}$ & $5.363\ten{-7}$ & \textbf{8.395e-8} & 0.1610          \\ \hhline{|=|=|=|=|=|=|=|=|=|=|=|=|=|}
\multirow{12}{*}{F}
 & $\mathcal{E}^{\mathrm{Wave}}_{L}$ (a)     
     & 3.4640          & 9.0242          & 7.7271          & 4.8407          & 1.3265          & \textbf{0}      & \textbf{0}      & \textbf{0}      & \textbf{0}      & $1.251\ten{-12}$ & \textbf{0}      \\ \cline{2-13}
 & $\mathcal{E}^{\mathrm{Wave}}_{E}$ (a)        
     & $7.255\ten{-2}$ & 0.2801          & 0.3540          & 0.9560          & 0.6759          & $2.892\ten{-2}$ & 0.9706          & 8.0592          & 8.0592          & $3.415\ten{-2}$ & \textbf{1.566e-3} \\ \cline{2-13}
 & $\mathcal{E}^{\mathrm{Wave}}_{\gamma}$ (a)
     & 18.2674         & 4.3616          & 9.8222          & 23.7379         & 30.9775         & 0.2598          & 6.4103          & 20.4328         & 20.4328         & 0.1711          & \textbf{4.099e-3} \\ \cline{2-13}
 & $\mathcal{E}^{\mathrm{Wave}}_{L}$  (b)     
     & $1.253\ten{-2}$ & $3.266\ten{-2}$ & $1.595\ten{-2}$ & $2.162\ten{-2}$ & $3.424\ten{-2}$ & \textbf{0}      & \textbf{0}      & \textbf{0}      & \textbf{0}      & $2.228\ten{-12}$ & \textbf{0}      \\ \cline{2-13}
 & $\mathcal{E}^{\mathrm{Wave}}_{E}$ (b)      
     & $1.235\ten{-2}$ & $3.101\ten{-2}$ & $9.212\ten{-3}$ & $1.259\ten{-2}$ & $2.590\ten{-2}$ & $3.165\ten{-3}$ & \textbf{4.657e-6} & 0.4541          & 0.4541          & $6.172\ten{-5}$ & $3.428\ten{-5}$ \\ \cline{2-13}
 & $\mathcal{E}^{\mathrm{Wave}}_{\gamma}$ (b)       
     & $1.363\ten{-2}$ & $3.206\ten{-2}$ & $1.048\ten{-2}$ & $1.243\ten{-2}$ & $2.568\ten{-2}$ & $2.212\ten{-2}$ & 0.1447          & 12.3695         & 12.3695         & $1.972\ten{-4}$ & \textbf{1.060e-4} \\ \cline{2-13}
 & $\mathcal{E}^{\mathrm{Wave}}_{E}$ (c)      
     & 0.5667          & 0.1241          & 0.7610          & 0.1215          & 2.7264          & $5.902\ten{-5}$ & $4.578\ten{-2}$ & 4.4362          & 4.4362          & $1.417\ten{-5}$ & \textbf{1.597e-6} \\ \cline{2-13}
 & $\mathcal{E}^{\mathrm{Wave}}_{\gamma}$ (c)      
     & 27.0589         & 9.6524          & 6.8514          & 472.8658        & 6.1320          & $4.265\ten{-3}$ & 4.0600          & 204.6991        & 204.6991        & $6.983\ten{-4}$ & \textbf{4.074-5} \\ \cline{2-13}
 & $\mathcal{E}^{\mathrm{Wave}}_{E}$ (d)      
     & $1.546\ten{-2}$ & $3.184\ten{-2}$ & $1.580\ten{-2}$ & $1.657\ten{-2}$ & $3.424\ten{-2}$ & $6.214\ten{-6}$ & $7.991\ten{-6}$ & $5.833\ten{-2}$ & $5.833\ten{-2}$ & $1.319\ten{-7}$ & \textbf{5.055e-8} \\ \cline{2-13}
 & $\mathcal{E}^{\mathrm{Wave}}_{\gamma}$ (d)      
     & $1.622\ten{-2}$ & $3.302\ten{-2}$ & $1.556\ten{-2}$ & $1.630\ten{-2}$ & $3.311\ten{-2}$ & $4.284\ten{-4}$ & 0.1505          & 13.4672         & 13.4672         & $2.626\ten{-6}$ & \textbf{1.606e-6} \\ \cline{2-13}
 & $\mathcal{E}^{\mathrm{Wave}}_{\gamma}$ (e)      
     & $1.596\ten{-2}$ & $3.299\ten{-2}$ & $1.575\ten{-2}$ & $1.648\ten{-2}$ & $3.328\ten{-2}$ & $7.458\ten{-7}$ & 0.1380          & 0.2860          & 0.2860          & $2.248\ten{-7}$ & \textbf{1.059e-8} \\ \cline{2-13}
 & $\mathcal{E}^{\mathrm{Wave}}_{\gamma}$ (f)      
     & $1.604\ten{-2}$ & $3.299\ten{-2}$ & $1.593\ten{-2}$ & $1.660\ten{-2}$ & $3.352\ten{-2}$ & $1.871\ten{-6}$ & 0.1380          & 0.2860          & 0.2860          & $3.721\ten{-3}$ & \textbf{5.596e-9} \\
     \hhline{|=|=|=|=|=|=|=|=|=|=|=|=|=|}
\multirow{6}{*}{G}
 & $\mathcal{E}^{\mathrm{Osc}}_{E}$ (a)      
     & \textbf{0}      & \textbf{0}      & \textbf{0}      & \textbf{0}      & \textbf{0}      & \textbf{0}      & \textbf{0}      & 2.0078          & 5.4507          & \textbf{0}      & \textbf{0}      \\ \cline{2-13}
 & $\mathcal{E}^{\mathrm{Osc}}_{\varphi}$ (a)  
     & 0.7117          & 0.5958          & 0.7051          & 0.6249          & 0.5801          & 0.3747          & \textbf{5.422e-2} & 3.1003          & 2.2347          & 2.5209          & 1.6672          \\ \cline{2-13}
 & $\mathcal{E}^{\mathrm{Osc}}_{u}$ (a)  
     & \textbf{0}      & \textbf{0}      & \textbf{0}      & \textbf{0}      & \textbf{0}      & \textbf{0}      & \textbf{0}       & 3.1513          & 7.0100          & \textbf{0}      & \textbf{0}      \\ \cline{2-13}
 & $\mathcal{E}^{\mathrm{Osc}}_{E}$ (b)   
     & \textbf{0}      & \textbf{0}      & \textbf{0}      & \textbf{0}      & \textbf{0}      & \textbf{0}      & \textbf{0}     & $8.885\ten{-2}$ & $4.230\ten{-13}$ & \textbf{0}      & \textbf{0}      \\ \cline{2-13}
 & $\mathcal{E}^{\mathrm{Osc}}_{\varphi}$ (b)  
     & $7.803\ten{-4}$ & $2.464\ten{-3}$ & $5.372\ten{-4}$ & $1.121\ten{-3}$ & $1.804\ten{-3}$ & $6.113\ten{-2}$ & \textbf{4.656e-4} & 0.4343          & $1.472\ten{-2}$ & $5.978\ten{-2}$ & $3.009\ten{-2}$ \\ \cline{2-13}
 & $\mathcal{E}^{\mathrm{Osc}}_{u}$ (b)  
     & \textbf{0}      & \textbf{0}      & \textbf{0}      & \textbf{0}      & \textbf{0}      & \textbf{0}      & \textbf{0}    & $3.489\ten{-2}$ & $1.010\ten{-6}$ & \textbf{0}      & \textbf{0}  \\
\hline
    \end{tabular}
    }
    \caption{Summary of errors of all schemes and various tests. The columns denote the schemes as described in table \ref{tab:schemes}. The rows are grouped according to the test cases described in the main text. The error measures are described in the main text. Unless otherwise stated, errors with a magnitude below $10^{-13}$ are considered numerical rounding errors and reported as zero. In each row, the smallest errors are typeset in boldface, highlighting the schemes with the highest accuracy in the corresponding test case.
    }
    \label{tab:error_measures}
\end{sidewaystable}

\subsection{\label{sec:gyromotion} Case A: Gyromotion with $\mathbf{E}=0$ and constant $\mathbf{B}$}

The gyromotion of a particle in the absence of an electric field, $\mathbf{E}=0$, with a constant magnetic field, $\mathbf{B} = B_z \mathbf{\hat{z}}$ allows the determination of phase and energy when compared with analytical results. A particle was initialised with $\mathbf{u}_0 = c(\gamma_0^2-1)^{1/2}\mathbf{\hat{x}}$ and propagated for 10 steps with the timestep given by $\Delta t \omega_c / 2\pi= 0.1$ where $\omega_c = q B /\gamma_0 m$ is the relativistic cyclotron frequency. The particles were initialised with $\gamma_0=1.001$ for the non-relativistic case (a) and  $\gamma_0=10$ for the relativistic case (b). Two error measures are calculated, 
\begin{displaymath}
    \mathcal{E}_\phi^{\mathrm{G}} = \arctan\left(\frac{u_y}{u_x}\right)
    \qquad\text{and}\qquad
    \mathcal{E}_u^{\mathrm{G}} = \frac{|u|}{|u_0|}.
\end{displaymath}
$\mathcal{E}_\phi^{\mathrm{G}}$ measures the error in the cyclotron frequency of a simulated particle whereas $\mathcal{E}_u^{\mathrm{G}}$ is a measure for the error in energy conservation. Both errors were evaluated at the end of the simulation when the particle had performed a full orbit.

The rows in table \ref{tab:error_measures} denoted by \textbf{A} show the errors for the gyromotion test. Note that, by construction, all schemes considered here are exactly energy-preserving in the absence of an electric field. This can easily be seen since, for $\mathbf{E}=0$ the velocity update in all schemes amounts to a simple rotation in velocity space. The phase error $\mathcal{E}_\phi$ obtained for Boris matches the well-known fact that the Boris scheme rotates the velocity by $\theta_B = 2 \arctan(\omega \Delta t/2)$ each time step, instead of the exact $\theta = \omega_c \Delta t$. In the absence of an electric field the midpoint schemes Vay, ZZ, and IMP are identical to Boris and show the same errors. For non-relativistic velocities, the phase error of HC is almost the same as that of the Boris scheme but changes sign for relativistic velocities. This implies that, for this choice of $\Delta t$ there is a Lorentz factor for which the phase error of HC vanishes. However, the phase error of HC converges to the finite value of $\mathcal{E}_\phi^{\mathrm{G}} \approx -0.1082$ as $\gamma\to\infty$ at a magnitude that is lower by almost a factor of 2 when compared to Boris. All other schemes have errors that are independent of $\gamma$. Chin \& Cator's CC method has a phase error with a magnitude that is reduced by almost a factor of two with respect to Boris. The phase error of CC has the opposite sign owing to the fact that the rotation angle is overestimated in CC whereas it is underestimated in the other type I schemes.
By construction, the phase errors of GYR, and the type II schemes, with the exception of ZZ, are identically zero in this test case. These schemes use trigonometric functions to perform the exact velocity rotation, $\omega_c \Delta t$. In contrast, the second-order scheme ZZ performs a second order velocity update similar to Boris and, therefore, exhibits a phase error comparable to the second-order schemes solved in the simulation frame of reference. The transformation into the particle's rest frame on its own does not provide any added numerical accuracy in this test case.

\subsection{\label{sec:lorentz_boost} Case B: Lorentz-boosted particle at rest, $\mathbf{E} = -\mathbf{v}_0\times\mathbf{B}$}

\begin{figure}
    \centering
    \includegraphics[width=0.75\textwidth]{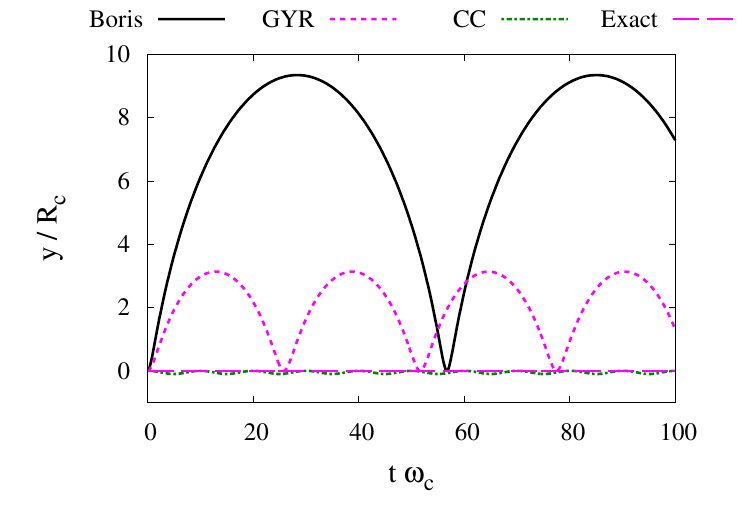}
    \caption{The $y$--coordinate of the particle trajectory vs time in a crossed electric and magnetic field with $\mathbf{E} = -\mathbf{v}_0\times\mathbf{B}$ (test case B) for a selection of integration schemes. Time is normalised by $\omega_c$ and the $y$--coordinate by $R_c = v_0/\omega_c$. The initial Lorentz-factor of the particles is $\gamma_0=10$. The curve labeled "Exact" denotes the exact solution which is reproduced by the pushers Vay, HC, PL, LiLF, GH, GH2, ZZ, and IMP.}
    \label{fig:lorentz_trajectory}
\end{figure}

Given a particle with initial velocity $\mathbf{v}_0 = \mathbf{u}_0/\gamma_0$, this test case investigates the trajectory in a crossed electric and magnetic field with $\mathbf{E} = -\mathbf{v}_0\times\mathbf{B}$. In this case, the forces from the electric and magnetic field acting on the particle exactly cancel and the particle is expected to propagate in a straight line with constant velocity $\mathbf{v}_0$. In the particle's rest frame, the electric field vanishes. Thus, this case can be interpreted as a particle at rest in a constant magnetic field viewed from a Lorentz-boosted frame moving with velocity $-\mathbf{v}_0$.
Many second-order numerical integrators treat the acceleration due to the electric and magnetic fields separately and not every scheme guarantees that these effects exactly cancel in this configuration. 

For the test presented here, the magnetic field pointed in the $z$--direction while the electric field was chosen to point in the $y$--direction. A particle was initialised with $\mathbf{u}_0 = c(\gamma_0^2-1)^{1/2}\mathbf{\hat{x}}$. The values of $\gamma_0$ were chosen to be (a) 1.001, (b) 10, (c) 100, (d) 1000, and (d) 10000. The time step was $\Delta t \omega_c / 2 \pi = 0.1$.

Figure \ref{fig:lorentz_trajectory} shows the trajectory of the $y$--coordinate over time for a selection of integration schemes when $\gamma_0=10$. The displacement from the exact path $y=0$ shows a periodic behaviour for all schemes, but the period of the path is different for each scheme. Schemes with larger errors also exhibit longer periodic behaviour. To ensure that the full error of each scheme was measured, the total simulation time for runs (a-c) was chosen such that at least one full period of the path with the largest error was completed. While the period of the path depends on the integration scheme, it was found to scale approximately with $\gamma_0^2$. This makes it infeasible to simulate the full period for the runs with largest values of $\gamma_0$. For runs (d,e) the simulation time was restricted to two cyclotron periods.

As a measure for the error in the propagation direction and the energy of the particle, two error measures are introduced,
\begin{displaymath}
    \mathcal{E}_\phi^{\mathrm{L}} = \max\left|\arctan\left(\frac{u_y}{u_x}\right)\right|
    \quad\text{and}\quad
    \mathcal{E}_\gamma^{\mathrm{L}} = \max\left| \frac{\gamma - \gamma_0}{\gamma_0 - 1}\right|.
\end{displaymath}
The maximum is taken along the full trajectory of the particle during the simulation. $\mathcal{E}_\phi^{\mathrm{L}}$ measures the maximum angle of the velocity away from its starting velocity in the $x$--direction, while $\mathcal{E}_\gamma^{\mathrm{L}}$ measures the maximum relative error of the kinetic energy $\gamma - 1$. This error in kinetic energy can also be taken as a measure of the trajectory deviation from a straight line as the particle gains energy by moving along the electric field in the $y$--direction. As can be expected, the maximum error of the kinetic energy is observed half-way through the periodic orbit. On the other hand, it was found that the maximum error in the propagation angle occurs early on during the cycle. This means that for the highest $\gamma_0$ runs, (d) and (e), where the full orbit isn't realised, $\mathcal{E}_\phi^{\mathrm{L}}$ is still a meaningful measure for the accuracy of the schemes.

The errors for the different schemes are tabulated in section B of Table \ref{tab:error_measures}. While Boris converges to the exact result in the non-relativistic case, $\gamma_0 = 1.001$, it stands out as one of the schemes with the largest errors in the relativistic case, both for $\mathcal{E}_\phi^{\mathrm{L}}$ and $\mathcal{E}_\gamma^{\mathrm{L}}$. CC and GYR show finite errors. The errors for GYR increase with $\gamma_0$ but are lower than Boris for the relativistic cases. Interestingly, the errors for CC, while being the largest for the non-relativistic case, shrink with increasing $\gamma_0$. Vay, HC, PL, LiLF, GH, GH2, ZZ, and IMP produce exact results up to machine precision for low $\gamma_0$. While Vay and HC have been designed to produce the correct trajectory specifically for this test case, PL, LiLF, GH and GH2 solve the exact trajectory for any constant field configuration. However, for large $\gamma_0$ some numerical error can be observed for all these schemes. The appearance of errors above the reporting threshold of $10^{-13}$ for larger values of $\gamma_0$ can be attributed to the accumulation of small rounding and root finding errors over many time steps. As $\gamma_0$ grows, so do the fields $\mathbf{E}$ and $\mathbf{B}$ and the calculation of the resulting force turns into a subtraction of large numbers with the corresponding loss of precision. Of the exact schemes, PL and LiLF are the ones that show accumulation of errors for the lowest values of $\gamma_0$. The main reason for this can be attributed to the finite tolerance in solving the implicit equation (\ref{eq_proper_time}) for the proper time step, $\Delta \tau$. It was found that decreasing the accuracy of the root solver resulted in larger errors here. Another reason for the errors may be due to the fact that, in the straightforward implementation of the scheme, the force term has to cancel out in multiple places, e.g. in the evaluation of $u_\sigma$ as well as in the term $F u_\kappa$ in equations (\ref{eq_exact_x}) and (\ref{eq_exact_u}). Small rounding errors could thus add up more quickly compared to other schemes. Notably, IMP produces exact results without such an accumulation of rounding errors even though it is only a second order scheme that was not specifically designed for this test case and the implicit solver uses a larger tolerance of $1e-13$ when compared to the tolerances used by PL and LiLF. The values of $\mathcal{E}_\phi^{\mathrm{L}}$ for the $\gamma_0 = 1000, 10000$ in cases (d) and (e) are consistent with the those for lower $\gamma_0$. 

\subsection{\label{sec:exb_drift} Case C: Gyration in a Lorentz-boosted frame}

\begin{figure}
    \centering
    \includegraphics[width=0.5\textwidth]{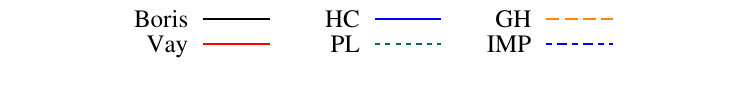}
    
    \includegraphics[width=0.49\textwidth]{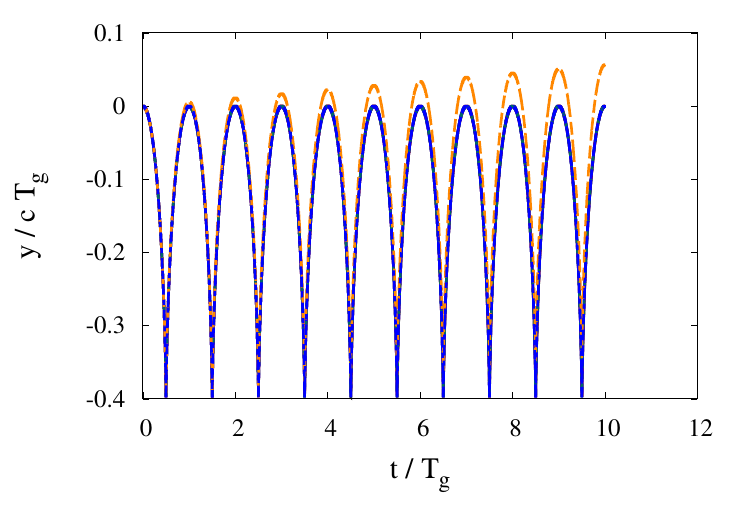}
    \includegraphics[width=0.49\textwidth]{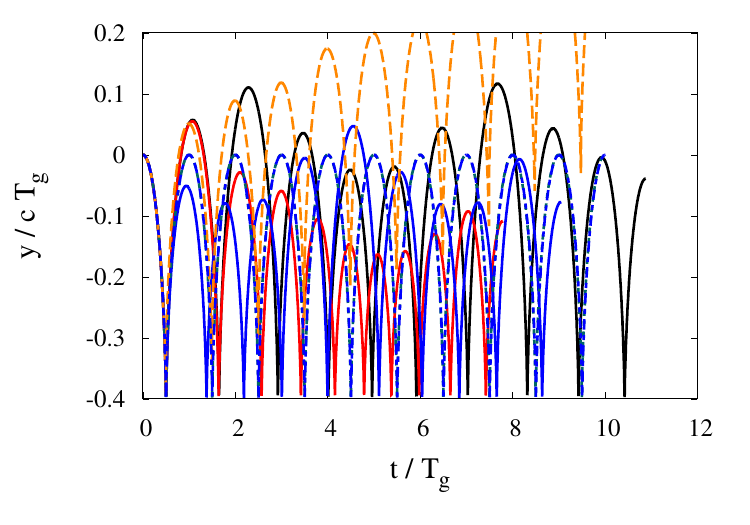}
    \caption{The $y$--coordinate of the particle trajectory orbiting in a crossed electric and magnetic field (test case C) for a selection of schemes. The time steps are $\Delta t = 4\times10^{-4} T_g$ (left) and $2\times10^{-3} T_g$ (right). For $\Delta t = 4\times10^{-4} T_g$ all curves, with the exception of GH, are indistinguishable. For $2\times10^{-3} T_g$ only the curves for PL and IMP show the correct behaviour and lie on top of each other. }
    \label{fig:exb_trajectory}
\end{figure}

The next test considers the gyration of a particle in a crossed electric and magnetic field, such that the forces do not cancel exactly. The particle motion consists of a gyration superimposed onto a drift in the $\mathbf{E}\times\mathbf{B}$ direction. Given $E < cB$, there exists a boosted frame $M$ with a boost velocity $v_M$ in which the electric field, $\mathbf{E}_M$, vanishes and the particle performs a pure circular motion in a magnetic field $\mathbf{B}_M$ with a velocity $v_P$ corresponding to a Lorentz factor $\gamma_P$. Thus, in the boosted frame, the cyclotron frequency for the relativistic particle becomes $\omega_{c,M} = qB_M/ (\gamma_P m)$. If the Lorenz factor of the boosted frame is denoted by $\gamma_M = (1 - v_M^2/c^2)^{-1/2}$, then the period of motion in the lab frame becomes $T_g  = 2\pi \gamma_P \gamma_M m / (q B_M)$. On the other hand, the magnetic field in the laboratory frame is boosted by $B_L = \gamma_M B_M$ which means that the cyclotron period for a non-relativistic particle evaluates to $T_{c,0}  = 2\pi m / q B_M \gamma_M = T_g / (\gamma_M^2 \gamma_P)$.

Simulations were performed of a particle initialized with a velocity in the $x$--direction given by $\gamma_P = 10, \gamma_M = 5$ for the mildly relativstic run (a) and $\gamma_P = 50, \gamma_M = 25$ for the strongly relativistic run (b). Note that in run (b) the maximum Lorentz factor of the particle in the simulation frame is $\gamma_P \gamma_M = 1250$. Figure \ref{fig:exb_trajectory} shows the $y$--coordinate of the orbits calculated for case (a) by a selection of integrators for two different time steps $\Delta t = 4\times10^{-4} T_g$ and $2\times10^{-3} T_g$. For the smaller time step, all integrators shown in the figure perform well within plotting accuracy, except for GH. GH is shown as an example of an integrator that fails to reproduce the exact trajectory even for the smaller time step. The other integrator that exhibit noticeable errors for $\Delta t = 4\times10^{-4} T_g$ is GH2 (not shown). For the larger time step, most solutions deviate substantially from the exact solution, even though each cycle is resolved by 500 time steps. The exception are the PL, LiLF, ZZ and IMP integrators. PL is exact by design for arbitrary time steps. One can observe that the main errors are introduced at the cusps of the trajectories. At the cusps, the particle motion in frame $M$ is directed in the negative $x$--direction with a larger velocity than the frame boost. As a result, the particle trajectory in the laboratory frame also possesses a negative $x$--velocity and loops back on itself. Note, that the chosen time steps correspond to $\Delta t = 0.1 T_{c,0}$ and $0.5 T_{c,0}$.  The resulting velocity in the laboratory frame is non-relativistic which implies that the non-relativistic time scale $T_{c,0}$ should be resolved.

To quantify the accuracy of the schemes for this test, the particle trajectory is integrated over $10 T_g$. The duration of the period of motion is evaluated by detecting the zero crossings of the velocity component $u_y$. Let the $y$--velocity component at a time step $t_n$ be denoted by $u_y^{(n)}$. A gyration period is completed when $u_y$ changes sign from positive to negative, $u_{y}^{(n)} > 0 \ge u_{y}^{(n+1)}$. To find an approximation to the $k$-th crossing time, $T_k$, a linear interpolation is used $T_k = t_{n+1} -  (u_y^{(n+1)}/(u_y^{(n+1)} - u_y^{(n)}))\Delta t$. A similar interpolation is used to find the $x$--position, $x_k$ at time $T_{k}$. Using this, the following error measures are introduced,
\begin{displaymath}
    \mathcal{E}_t^{\mathrm{ExB}} = \frac{T_{10} - 10 T_g}{10 T_g}
    \quad\text{and}\quad
    \mathcal{E}_x^{\mathrm{ExB}} = \frac{x_{10} - T_{10} v_M}{T_{10} v_M}.
\end{displaymath}
Part C in table \ref{tab:error_measures} shows the errors $\mathcal{E}_t^{\mathrm{ExB}}$ and $\mathcal{E}_x^{\mathrm{ExB}}$ for the different schemes for $\Delta t = 0.1 T_{c,0}$. All schemes show small values for both errors. As stated above, PL is exact by construction but still shows an accumulation of errors above the reporting threshold that can be attributed to the finite accuracy of the solution for the proper time step $\Delta \tau$. LiLF shows a comparable error in the gyration period, but a larger error for the distance traveled. Interestingly, Vay, PL and ZZ all exhibit near zero error for $\mathcal{E}_x^{\mathrm{ExB}}$, but Vay and ZZ shows a non-negligible error for $\mathcal{E}_t^{\mathrm{ExB}}$.

In the highly relativistic case (b) the errors for Vay, HC, GYR, CC, GH, and GH2 increase substantially. The remaining schemes also show an increase in both error measures but this increase is more modest. PL performs best with errors that can be attributed to root-finding and rounding errors only.

\subsection{Case D: Parallel fields with spatial variation \label{sec:hg_test}}

Higuera \& Cary \cite{Higuera:2017} proposed a test in which the electric and magnetic potentials have the form $\Phi=ax^2/2$ and $\mathbf{A}=by^2\mathbf{\hat{z}}$, resulting in the fields $q\mathbf{E}/mc^2 =-a x\mathbf{\hat{x}}$ and $q\mathbf{B}/mc =by\mathbf{\hat{x}}$. Here, the particle oscillates in the $x$--direction in the harmonic electrostatic potential while simultaneously gyrating in the $y$--$z$ plane. In the non-relativistic limit, the period of oscillation in the $x$--direction is given by $T_{\text{osc}} = 2\pi / \sqrt{a}$. The Hamiltonian
\begin{displaymath}
    H = c\sqrt{m^2c^2 + p_x^2 + p_y^2 + (p_z - qA_z(y))^2} + q\Phi(x)
\end{displaymath}
is conserved, as is the canonical momentum $p_z$ and the invariant
\begin{equation}
    I_y = p_y^2 + (p_z - qA_z(y))^2.\label{eq:hg_invariant_full}
\end{equation}
Here $p_x$, $p_y$, and $p_z$ are the components of the canonical momentum $\mathbf{p} = m\mathbf{u} + q\mathbf{A}$. If the trajectories are chosen such that $p_z(t=0)=0$ then $p_z$ will remain zero at all times and the invariant can be written as
\begin{equation}
    I_y = p_y^2 + q b y^4 = p_{y,0}^2,\label{eq:hg_invariant}
\end{equation}
where $p_{y,0}$ is the $y$-component of the momentum at $y=0$. In \cite{Higuera:2017} Poincar\'e plots are used to qualitatively judge the performance of their integration scheme. These plots are generated by recording $y$ and $p_y$ at the intersections of the trajectory with the $x=0$ plane. For each set of initial parameters, the resulting points should lie on the curve in the $y$--$p_y$ plane given by these parameters and equation (\ref{eq:hg_invariant}). To allow comparison with their results the same parameters, $a=1$ and $b=2$, are chosen here. The particle is initialised with $x=y=z=p_z=0$ and varying values of $p_y$. The momentum $p_x$ is then chosen such that the total energy matches the prescribed, mildly relativistic value of $H=4mc^2$. The test was designed to expose the effects of any varying phase space volume elements introduced by the schemes. These effects are strongest in the mildly relativistic case which justifies this specific choice of parameters \cite{Higuera:2017}. 

\begin{figure}
    \centering
    \includegraphics[width=0.49\textwidth]{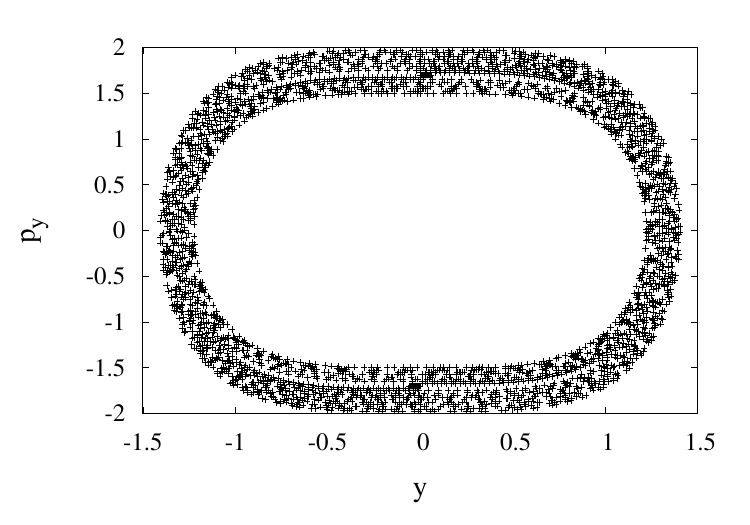}
    \includegraphics[width=0.49\textwidth]{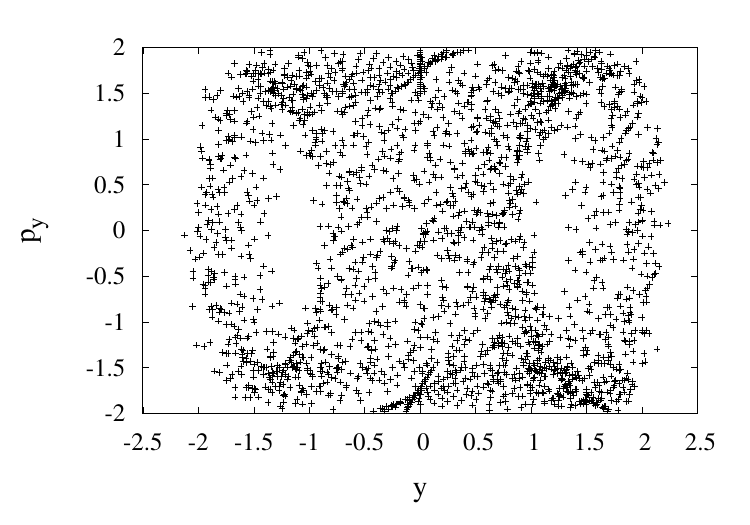}
    \caption{Poincar\'e plots for the particle orbits in parallel $\mathbf{E}$ and $\mathbf{B}$ fields with spatial variation, described in section \ref{sec:hg_test}. The time step is $\Delta t = T_{\text{osc}}/40$. The remaining parameters are given in the main text. Only the results for the Boris scheme (left) and the PL scheme (right) are shown.}
    \label{fig:hg_poincare}
\end{figure}

Plot \ref{fig:hg_poincare} shows the resulting Poincar\'e plots for two chosen integrators and a time step of $\Delta t = T_{\text{osc}}/40$. The Boris scheme is shown in the left panel. It shows that the orbits remain on the curves given by equation (\ref{eq:hg_invariant}) for the chosen parameters. In contrast, the PL integration scheme produces a plot, shown in the right panel, in which the structure has completely changed and the original curves cannot be recognised. All other integrators produce plots that are more similar that to the Boris scheme than to PL for this choice time step.

To obtain quantitative error measures, the particle orbit was integrated for 10 oscillation periods with a starting momentum of $p_y = 1.7mc$. This value is located near the resonance defined by the starting conditions at which the relativistically corrected oscillation period in the electric field coincides with the period of gyration. Three errors are calculated,
\begin{gather*}
    \mathcal{E}^{\mathrm{HC}}_{H} = \max\left|\frac{H - H_0}{H_0}\right|,
    \quad
    \mathcal{E}^{\mathrm{HC}}_{I} = \max\left|\frac{I_y - I_0}{(mc)^2}\right|,\\
    \text{and}\quad
    \mathcal{E}^{\mathrm{HC}}_{p} = \max\left|\frac{p_z}{mc}\right|,
\end{gather*}
with $I_0 = p_{y,0}^2$ and $I_y$ is the full form of the invariant given by equation (\ref{eq:hg_invariant_full}) The maximum is taken along the full trajectory, not just at the intersection points with the $x=0$ plane. The resulting errors are tabulated in section D of Table \ref{tab:error_measures} for a time step of $\Delta t = 0.1 T_{\text{osc}}$. The qualitative results from the Poincar\'e plot appear to suggest a violation of the invariant $I_y$. However, equation (\ref{eq:hg_invariant}) for the expected shape of the plot was derived under the assumption that $p_z$ remains zero at all times. The results in table \ref{tab:error_measures} show that almost all integration schemes conserve $I_y$ up to machine precision. The only exception to this rule is the Vay scheme which is also the only scheme in this comparison that is not volume preserving \cite{Higuera:2017}. For the remaining schemes, performance is down to their ability to conserve the canonical momentum $p_z$. Here PL, GH, and GH2 stand out as having the largest errors while type I schemes perform overall better. The implicit solver IMP is able to conserve all invariants to machine precision.

\begin{figure}
    \centering
    \includegraphics[width=0.75\textwidth]{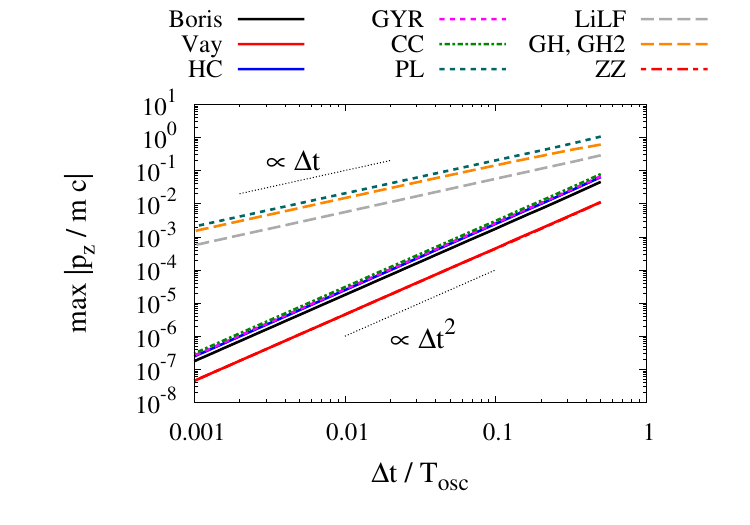}
    \caption{Scaling of the numerical error in the canonical momentum $p_z$ with the time step for the particle orbits in parallel $\mathbf{E}$ and $\mathbf{B}$ fields with spatial variation, described in section \ref{sec:hg_test}.
    IMP is not shown because the error is zero within machine precision. GH and GH2 produce identical results. GYR lies almost on top of HC with a slight shift less than the line width. Similarly, Vay and ZZ lie on top of each other.
    The dotted lines indicate ideal scaling with $\Delta t$ and $\Delta t^2$.
    }
    \label{fig:hg_scaling}
\end{figure}

In order to shed light on the poor performance of PL, GH, and GH2 in this test, the convergence of the $p_z$ error with the time step has been investigated. Figure \ref{fig:hg_scaling} shows the scaling of this error with $\Delta t$. Surprisingly, PL, LiLF, GH, and GH2, i.e. all type II schemes with the exception of ZZ, exhibit only first order convergence while all other explicit integrators show second order convergence as expected. In the PL scheme, $\mathbf{x}$ and $\mathbf{u}$ are both given at the full time step $t_n$. The evaluation of the fields and the proper time step $\Delta \tau$ is carried out at $t_n$ and $x_n$ and not at $t_{n+1/2}$ and $x_{n+1/2}$ as would be required for a second order scheme. But even for the schemes that stagger $\mathbf{x}$ and $\mathbf{u}$ these integration schemes have to invert the equation for $\Delta t$ in terms of the proper time $\Delta \tau$. This equation involves the dot product $\mathbf{E} \cdot \mathbf{u}$. For LiLF, GH, and GH2, these two quantities are not given at the same time but staggered, again reducing the order of the schemes for spatially varying fields. Although being a type II scheme, ZZ performs a leapfrog update of the position and velocity, resembling type I schemes with the difference that the update is performed using the proper time $\tau$. This splitting allows the fields to be evaluated at the half time step, $\Delta \tau/2$, which provides the second order convergence.

\subsection{Case E: Magnetic Bottle\label{sec:magnetic_bottle}}
% Ripperda's dt = 1e-7 corresponds to 2000\pi time-steps per fast orbit for gamma=100

The motion of a particle in a magnetic bottle, a configuration in which a particle is trapped between two magnetic mirrors, is commonly used to evaluate the long-term stability of particle integration schemes, see e.g. \cite{Ripperda:2018,Tretiak:2019,Ricketson:2020}. The motion of the particle consists of a fast gyration combined with a slow bounce between the two magnetic mirrors. Due to the difference of time-scales, the magnetic moment is an adiabatic invariant, meaning that it is approximately conserved up to second order in the ratio of the two scales. Here, the setup proposed by Ripperda et al \cite{Ripperda:2018} is used. The electric field is zero and the magnetic field is given in cylindrical coordinates as
\begin{displaymath}
    \mathbf{B}_{\text{Bottle}} = B_0\left[\left(1 + \frac{z^2}{L^2}\right)\mathbf{\hat{z}} - \frac{rz}{L^2}\mathbf{\hat{r}}\right],
\end{displaymath}
where $\mathbf{\hat{r}}$ is the radial unit vector perpendicular to $\mathbf{\hat{z}}$, and $r$ is the radial distance from the $z$--axis. $L$ is the length scale determining the size of the bottle.
The fast gyration frequency for a particle in the $z=0$ plane moving with a Lorentz factor of $\gamma_0$ is given by $\omega_c = qB_0/\gamma_0 m$. If the initial particle $\mathbf{u}$ is split into $\mathbf{u}_z$ and $\mathbf{u}_{\phi}$, parallel and perpendicular to the $z$--axis, then the Larmor radius of the particle in the $z=0$ plane is given by $r_L = m u_{\phi} / q B_0$. For the test, an aspect ratio of $L/r_L = 500$ is chosen and the particle is initialised with $\gamma_0=100$ and a pitch angle of 45$^\circ$, such that $u_x = u_{\phi} = u_z$ and $u_y=0$. The starting location of the particle is on the $y$--axis at $y=-r_L$. This ensures the centre of the gyromotion is located on the $z$--axis. The magnetic moment of the particle is given by $\mu_B = \gamma u_{\perp}^2 / |\mathbf{B}(\mathbf{x})|$. The time step is chosen to be $\Delta t = 0.1 T_c = 2\pi/10 \omega_c$. 

The quantities
\begin{displaymath}
    \mathcal{E}^{\text{Mirror}}_\gamma = \max\left| \frac{\gamma - \gamma_0}{\gamma_0}\right|,
    \quad\text{and}\quad
    \mathcal{E}^{\text{Mirror}}_\mu = \max\left| \frac{\mu_B - \mu_0}{\mu_0}\right|
\end{displaymath}
measure the errors in total energy and magnetic moment. Here $\mu_0 = \mu_B(t=0)$ is the magnetic moment at the start of the simulation. 

The results are summarised in section E of table \ref{tab:error_measures}. All integrations schemes, except Vay are found to be energy preserving. The Vay scheme exhibits a cumulative energy error over the duration of the ten bounces of $1.390 \times 10^{-3}$. The magnetic moment is not expected to be conserved exactly as it is only an adiabatic invariant, and its relative variation is expected to scale as $(r_L / L)^2 = 4\times 10^{-6}$. Most of the schemes exhibit errors below this value. Surprisingly, the IMP scheme has the largest error in this comparison. It should be noted that, in \cite{Ripperda:2018}, the test results presented for the IMP scheme used time steps of $\Delta t \le 1/100 \omega_c$ resulting in over 600 time steps for each fast orbit. The large error in the PL scheme is expected because it reduces to first order in the case of a non-constant magnetic field as seen in the previous test. The errors of LiLF, GH, GH2, and ZZ are compatible with the expected variations of the magnetic moment.

\subsection{\label{sec:rel_laser}Case F: Motion in a plane wave with relativistic intensity}

The particle motion in an electromagnetic plane wave was used in \cite{Arefiev:2015} as a test for the Boris integrator. In this setup, a wave is propagating in the $x$--direction with a wavelength $\lambda_0$ and polarised in the $y$--direction. Then $T_0 = \lambda_0 / c$ and the motion of the particle is governed by two integrals of motion \cite{Landau:1975},
\begin{displaymath}
    \frac{u_y}{c} + a = \text{const}
    \quad\text{and}\quad
    \gamma - \frac{u_x}{c} = \text{const},
\end{displaymath}
where $a = qA/mc$ is the normalised vector potential. Assuming the particle is at rest when $a=0$, one finds that
\begin{displaymath}
    \frac{u_y}{c} = -a
    \quad\text{and}\quad
    \frac{u_x}{c} = \frac{a^2}{2},
\end{displaymath}
and the maximum energy gained by the particle is given by \cite{Arefiev:2015}
\begin{equation}
    \gamma^* = 1 + \frac{a_0^2}{2},\label{eq:pw_gamma_star}
\end{equation}
where $a_0$ is the peak amplitude of the wave. 

\begin{figure}
    \centering
    \includegraphics[width=0.5\textwidth]{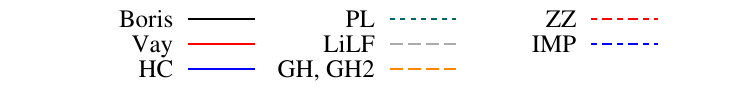}
    
    \includegraphics[width=0.49\textwidth]{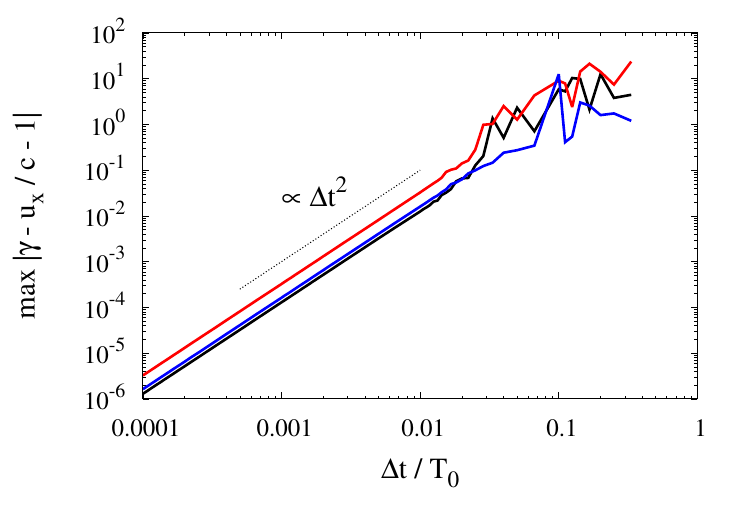}
    \includegraphics[width=0.49\textwidth]{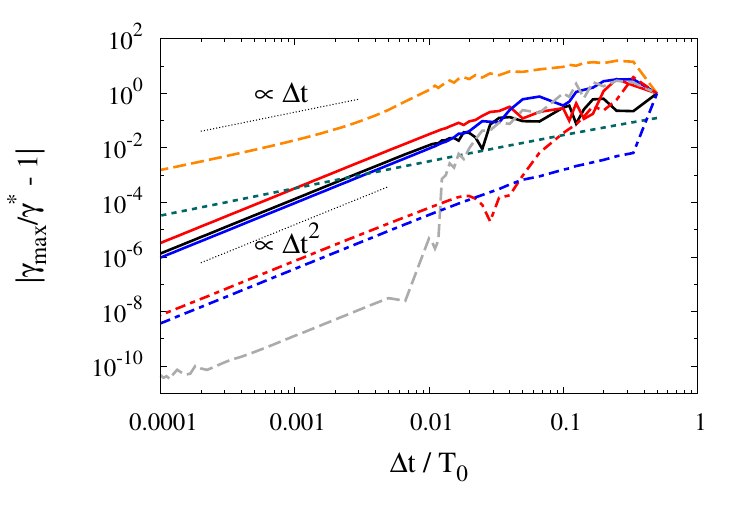}
    
    \caption{Numerical errors $\mathcal{E}^{\mathrm{Wave}}_{L}$ (left) and $\mathcal{E}^{\mathrm{Wave}}_{E}$ (right) of the calculated motion in a relativistic plane wave (test case F) as a function of the time step $\Delta t$. $\mathcal{E}^{\mathrm{Wave}}_{L}$ is within machine precision for PL, LiLF, GH, GH2, ZZ, and IMP. $\mathcal{E}^{\mathrm{Wave}}_{E}$ curves for CC and GYR are not shown because they are similar to HC and Vay respectively. The dotted lines indicate ideal scaling with $\Delta t$ and $\Delta t^2$.}
    \label{fig:planewave_scandt}
\end{figure}

Integrations of the particle motion were performed for $a_0 = 10, 100, 1000$, and $10000$. The particle is initialized at rest when $a=0$. The particle trajectory is integrated over a time period of $200T_0$, exept for the run with $a_0 = 100$ where it is integrated over $2\times10^{4} T_0$. For the run with $a_0 = 10$, convergence tests were performed by varying the time step from $\Delta t = 10^{-4}T_0$ to $0.3T_0$. Three errors are evaluated,
\begin{gather*}
    \mathcal{E}^{\mathrm{Wave}}_{L} = \max\left(\gamma - \frac{u_x}{c} - 1\right),
    \qquad
    \mathcal{E}^{\mathrm{Wave}}_{E} = \frac{\gamma_{\mathrm{max}}}{\gamma^*} - 1,\\
    \text{and}\qquad
    \mathcal{E}^{\mathrm{Wave}}_{\gamma} = \max\left(\frac{\gamma - 1 - a^2 / 2}{\gamma}\right),
\end{gather*}

where $\gamma_{\mathrm{max}}$ is the maximum Lorentz factor of the particle, and $a$ in the expression for $\mathcal{E}^{\mathrm{Wave}}_{\gamma}$ is evaluated at the particle position. 
Figure \ref{fig:planewave_scandt} shows plots for $a_0 = 10$ of the errors $\mathcal{E}^{\mathrm{Wave}}_{L}$ and $\mathcal{E}^{\mathrm{Wave}}_{E}$ for different time steps. Values of $\mathcal{E}^{\mathrm{Wave}}_{L}$ are within machine accuracy for PL, LiLF, GH, GH2, ZZ, and IMP and are not shown. All type I schemes show similar results and a second order convergence of the error. From this one could conclude that the the type II schemes which employ the exact solution in proper time are superior for the plane-wave scenario. However, when looking at the error in maximum particle energy, the roles are partially reversed. GH and GH2 produce indistinguishable results with a roughly first order convergence. PL also exhibits first order convergence, albeit with a substantially smaller coefficient. The poor performance of these type II schemes can again be attributed to the fact that these schemes revert to first order due to temporal alignment issues when evaluating the proper time step $\Delta \tau$, as explained in section \ref{sec:hg_test}. Gordon \& Hafizi \cite{Gordon:2021} show that the GH scheme is exact up to machine precision when evaluated in the proper time of the particle. This holds, by extension, also for the PL scheme which only differs from GH in the way the proper time step is related to the simulation frame time step. Thus, these type II schemes can be useful integration schemes in cases where no time step restrictions apply. 
In contrast, the energy error produced by ZZ is almost two orders of magnitude lower than those of the type I schemes, comparable to the IMP scheme. LiLF has errors comparable to the type I schemes for large time steps but drops another two orders of magnitude below ZZ for small time steps.

\begin{figure}
    \centering
    \includegraphics[width=0.49\textwidth]{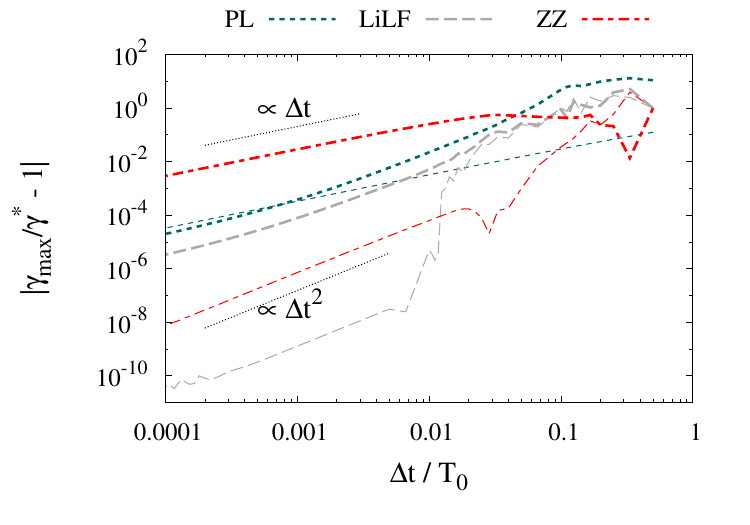}
    
    \caption{Numerical errors $\mathcal{E}^{\mathrm{Wave}}_{E}$ of the calculated motion in a relativistic plane wave (test case F) as a function of the time step $\Delta t$. Shown are results of the type II schemes that require numerical root solving. Thin lines are results obtained using SciPy's \texttt{brentq} function with \texttt{xtol=1e-30} and \texttt{rtol=1e-15}. Thick lines are obtained using the \texttt{newton} function with default parameters. The dotted lines indicate ideal scaling with $\Delta t$ and $\Delta t^2$.}
    \label{fig:planewave_scandt_prec}
\end{figure}

The low errors produced by ZZ and LiLF are attached with a caveat. It was found that the results are extremely sensitive to the root finding algorithm used in solving the implicit equation for the proper time. Figure \ref{fig:planewave_scandt_prec} shows a comparison of the $\mathcal{E}^{\mathrm{Wave}}_{E}$ for PL, LiLF, and ZZ using two different root finding algorithms. When using a standard Newton-Raphson method with the default parameters supplied by SciPy, the results degrade substantially.

A feature to note in figure \ref{fig:planewave_scandt} is the region with time steps larger than $\approx 0.05T_0$. Here, most schemes do not follow a clear scaling but instead show erratic dependence as a function of the time step. One should conclude from this that an individual error value for a specific time step in this range can not be taken as a true performance measure. Indeed, it was found that the exact numerical values of these errors are not reproducible when only slight changes to the implementation of the algorithms were made. This indicates that the relevant physical time scale is not resolved and small rounding errors amplify uncontrollably. With this caveat, HC appears to perform better for $\mathcal{E}^{\mathrm{Wave}}_{L}$ when looking at the overall behaviour of the error in this region, while for $\mathcal{E}^{\mathrm{Wave}}_{E}$ no clear winner among the explicit schemes can be identified. 

In \cite{Arefiev:2015}, the large errors in this region for the Boris scheme were identified with the fact that, for large $a_0$ the time step may not resolve the gyrofrequency near the field maxima of the wave. The Boris scheme has a systematic error for large rotation angles $\Delta t \omega_C$, and thus it sounds reasonable to assume that these errors in the gyrophase are responsible for the errors in the constants of motion. However, the fact that the GYR scheme exhibits errors on the same magnitude indicates that this interpretation cannot be the whole truth because GYR does not have any systematic error in the gyrophase. Here a different interpretation of the error is proposed.
Since in a plane wave field, $E = cB$, and estimating the change in velocity $m \Delta u = q E \Delta t$, which is true in the absence of a magnetic field, one can easily derive the relation
\begin{displaymath}
    \omega_c \Delta t = \frac{qB}{m} \Delta t = \frac{qE}{mc} \Delta t = \frac{\Delta u}{c}.
\end{displaymath}
Thus, large changes in phase also correspond to large relativistic accelerations during one time step, potentially leading to substantial changes in the Lorentz factor $\gamma$ within a time step. Through $\gamma$, the gyrofrequency depends on $u$ in a non-linear way. Thus, the choice of how to average $\gamma$ in the evaluation of the Lorentz force, can have a large impact on the performance of the numerical scheme. Simply evaluating the exact gyrophase does not compensate for a large time step.

Section F in table \ref{tab:error_measures} summarises the results for this test and quantifies the errors for runs (a) $a_0 = 10$, $\Delta t = 0.1T_0$, (b) $a_0 = 10$, $\Delta t = 0.01T_0$, (c) $a_0 = 100$, $\Delta t = 0.01T_0$, (d) $a_0 = 100$, $\Delta t = 0.001T_0$, (e) $a_0 = 1000$, $\Delta t = 10^{-4}T_0$, and (f) $a_0 = 10^4$, $\Delta t = 10^{-5}T_0$. For run (a), the time step does not resolve the physical time scale and the errors of all schemes are large. Reducing the time step in run (b) shows the differences in the schemes more clearly with IMP having the lowest errors and GH, and GH2 the largest.  Runs (c) and (d) were performed with $a_0 = 100$ and two different time steps $\Delta t = 0.01T_0$ in run (c), and $0.001T_0$ in run (d). Run (c), despite having a time step equal to run (b) again underresolves the physical time scale and all errors are large with the exception of IMP and possibly LiLF. For the smaller time step in run (d) again IMP shows the lowest errors followed by LiLF. Extending the tests to higher $a_0$ requires the time step to be scaled with $a_0^{-1}$ in order to resolve the dynamics of the particle in the strong fields. On the other hand, the dephasing time of the particle in the simulation frame scales with $a_0^2$. This makes it infeasible to evaluate the error $\mathcal{E}^{\mathrm{Wave}}_{L}$ and $\mathcal{E}^{\mathrm{Wave}}_{E}$ which assume their maximum on the time scale of the dephasing time. However, it was found that the maximum relative error of the instantaneous $\gamma$ captured by $\mathcal{E}^{\mathrm{Wave}}_{\gamma}$ is assumed early on in the cycle. Thus, for the runs (e) and (f) the simulation time was arbitrarily capped at $200 T_0$ and only the values for $\mathcal{E}^{\mathrm{Wave}}_{\gamma}$ are reported. Continuing the time step scaling of runs (b) and (d), runs (e) and (f) test the schemes for the extreme relativistic case. For most schemes, the error $\mathcal{E}^{\mathrm{Wave}}_{\gamma}$ remains almost constant between these four runs. The exception are IMP and LiLF which both exhibit a reduction of the error with increasing $a_0$. However, while the error of IMP for $a_0 = 10^4$ is below $10^{-8}$, the error for LiLF levels off for runs (e) and (f) at a value of $0.2860$. It should be noted that, at these extreme intensities quantum mechanical effects can become important. Li et al \cite{Li:2021} extend the PL and LiLF schemes to include radiation reaction effects on the particle orbit.

\subsection{Case G: Oscillating electric field $\mathbf{E} \| \mathbf{B}$ \label{sec:oscillating_ez}}

The plane wave test in the previous section has the properties that $\mathbf{E} \perp \mathbf{B}$ and $|\mathbf{E}| = c |\mathbf{B}|$ at all times. This pure vacuum field is a special case in which the trajectories have a distinct shape. This is apparent in the fact that the type II schemes PL, LiLF, GH, and GH2 are treating these conditions as a special case in which the general scheme is replaced by a special solution.

In order to compare the general formulation of the schemes in a time-dependent scenario, a test is proposed in which a particle gyrates in a constant magnetic field $\mathbf{B} = B_0 \mathbf{\hat{z}}$ with a superimposed oscillating electric field $\mathbf{E} = E_0 \mathbf{\hat{z}} \cos(\omega_0 t)$. The motion can be integrated exactly, yielding
\begin{align*}
    u_x &= u_{\perp} \cos(\varphi(t)),\\
    u_y &= - u_{\perp} \sin(\varphi(t)),\\
    u_z &= \frac{qE_0}{m\omega_0} \sin(\omega_0 t),\\
    \frac{d\varphi}{dt} &= \frac{\omega_\perp}{\sqrt{1 + \eta^2 \sin(\omega_0 t)}}.
\end{align*}
Here $\eta = \frac{E_0}{c B_0}\omega_\perp$, $\omega_\perp = q B_0/\gamma_\perp m$, and $\gamma_\perp = \sqrt{1 + u_{\perp}^2/c^2}$. The equation for $\varphi(t)$ can be integrated exactly using elliptic integrals.

The parameters for the test are chosen such that the initial perpendicular velocity is mildly relativistic with $\gamma_\perp = 1.1$. The oscillation frequency is chosen to be $\omega_0 = \omega_\perp / 2$ and the electric field strength is $E_0 = 10 m \omega_0 c / q$ so that the maximum longitudinal velocity is strongly relativistic. The particle orbit is integrated for five electric field oscillations. At the end of the simulation, the particle is expected to have zero longitudinal velocity $u_z(t=5T_0)=0$ and unchanged total energy $\gamma(t=5T_0) = \gamma_\perp$. Three error measures are used to quantify the performance of the schemes.
\begin{gather*}
    \mathcal{E}^{\mathrm{Osc}}_{u} = \frac{u_z(t=5T_0)}{c},
    \quad
    \mathcal{E}^{\mathrm{Osc}}_{E} = \frac{\gamma(t=5T_0)}{\gamma_\perp} - 1,\\
    \text{and}\quad
    \mathcal{E}^{\mathrm{Osc}}_{\varphi} = \max \left| \varphi - \varphi_\text{exact}\right|,
\end{gather*}
where $\varphi$ is the calculated phase of the particle and $\varphi_\text{exact}$ is the exact phase.

\begin{figure}
    \centering
    \includegraphics[width=0.5\textwidth]{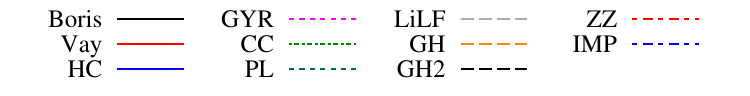}
    
    \includegraphics[width=0.49\textwidth]{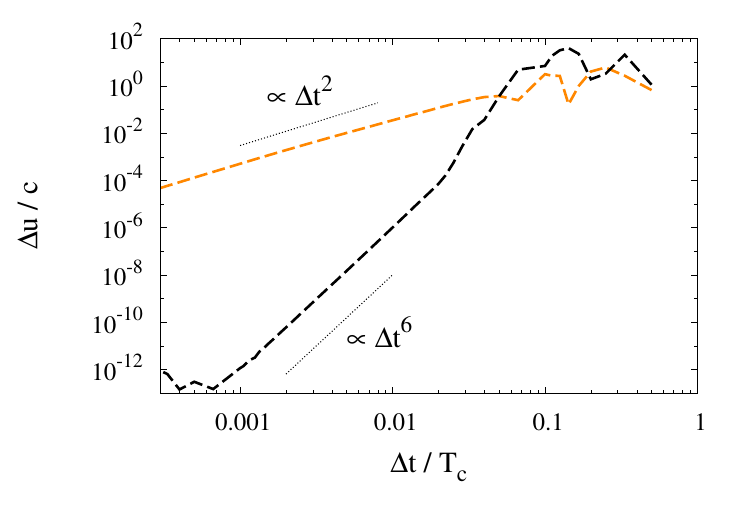}
    \includegraphics[width=0.49\textwidth]{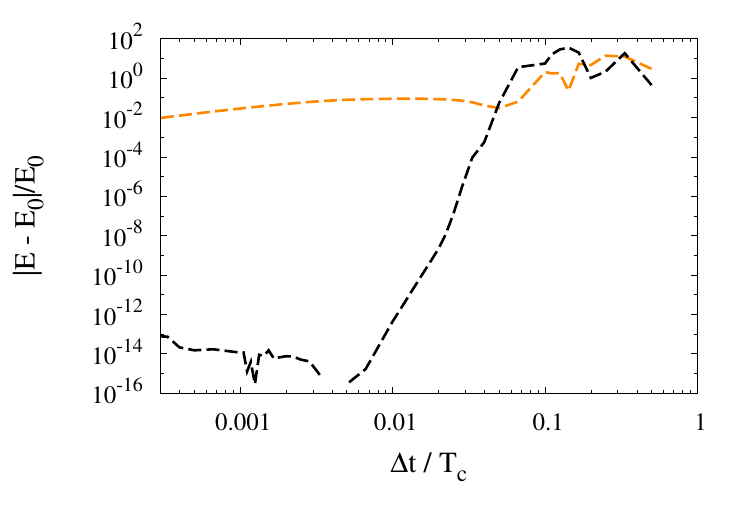}
    \includegraphics[width=0.49\textwidth]{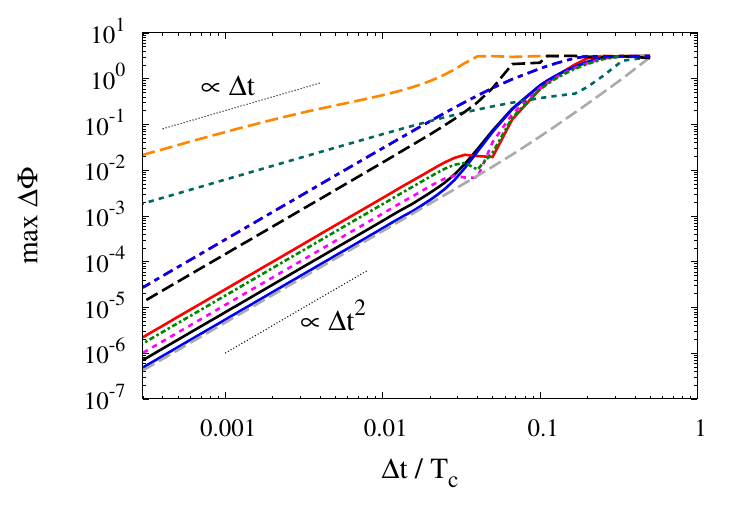}
    
    \caption{Numerical errors $\mathcal{E}^{\mathrm{Osc}}_{u}$ (top-left), $\mathcal{E}^{\mathrm{Osc}}_{E}$ (top-right) and $\mathcal{E}^{\mathrm{Osc}}_{\varphi}$ (bottom) of the calculated motion in a in an oscillating electric field, $\mathbf{E} \| \mathbf{B}$ (test case G), as a function of the time step $\Delta t$. $\mathcal{E}^{\mathrm{Osc}}_{u}$ and $\mathcal{E}^{\mathrm{Osc}}_{E}$ are within machine accuracy for all schemes except PL, LiLF, GH, and GH2. In the bottom plot, ZZ is not visible because it coincides with IMP within drawing accuracy.
     The dotted lines indicate ideal scaling with $\Delta t$, $\Delta t^2$, and $\Delta t^6$.
    \label{fig:oscillatingez_scandt}}
\end{figure}

Section G in table \ref{tab:error_measures} summarises the results for case (a) $\Delta t = 0.1 T_\perp$ and case (b) $\Delta t = 0.01 T_\perp$, where $T_\perp = 2\pi/\omega_\perp$. All schemes except PL, LiLF, GH, and GH2 conserve energy exactly and show no error in the parallel velocity $u_z(t=5T_0)$. For the larger time step CC has the smallest phase error, although this should not be taken as an overall indication of the performance of CC for this test. The convergence plots for the three error measures are shown in figure \ref{fig:oscillatingez_scandt}. For the measures $\mathcal{E}^{\mathrm{Osc}}_{u}$ and $\mathcal{E}^{\mathrm{Osc}}_{E}$ the four schemes PL, LiLF, GH, and GH2 are shown. GH, PL and LiLG appear to have second order convergence in $\mathcal{E}^{\mathrm{Osc}}_{u}$ for small time steps, whereas GH2 appears to have sixth order convergence. For $\mathcal{E}^{\mathrm{Osc}}_{E}$, GH shows only little improvement with decreasing time step, while GH2 and LiLF quickly drop to machine precision with decreasing time step and no clear convergence scaling can be observed. The scaling of $\mathcal{E}^{\mathrm{Osc}}_{E}$ for PL is approximately fourth order. The error measure $\mathcal{E}^{\mathrm{Osc}}_{\varphi}$ is non-zero for all schemes considered. For small time steps HC performs best. All type I schemes, IMP, LiLF, GH2 and ZZ exhibit second order convergence, whereas GH and PL only show first order convergence. LiLF performs best in this measure.

\section{\label{sec:higher_order} Extension to higher orders}

Yoshida \cite{Yoshida:1990} proposed a method for constructing higher order schemes from second order schemes for the case that the second order scheme can be written as a symplectic map over the Hamiltonian phase space. Let
\begin{displaymath}
    \Phi^n_{\Delta t} = \exp\left(\Delta t D_H^n\right)
\end{displaymath}
be an $n$-th order exponential map associated with the Hamiltonian vector field $D_H^n$. Here the superscript $n$ indicates the order of the scheme. Yoshida \cite{Yoshida:1990} proposes the following method. Given a scheme of order $n$, a new scheme of order $n+2$ can be constructed using the following prescription,
\begin{equation}
\begin{split}    
    \Phi^{n+2}_{\Delta t}
        &= e^{\alpha_1 \Delta t D_H^n} e^{\alpha_0 \Delta t D_H^n}e^{\alpha_1 \Delta t D_H^n}\\
        &=: e^{\Delta t D_H^{n+2}},\label{eq:yoshida_scheme}
\end{split}
\end{equation}
with
\begin{equation}
    \alpha_0 = -\frac{2^{1/(n+1)}}{2-2^{1/2(n+1)}},\qquad
    \alpha_1 = \frac{1}{2-2^{1/(n+1)}}.\label{eq:yoshida_coeff}
\end{equation}
The new update map maintains the symplectic structure because it is constructed as a sequence of symplectic maps. The derivation of the coefficients $\alpha_0$ and $\alpha_1$ and the proof of the order of the new scheme only makes use of the exponential map together with the Baker-Campbell-Hausdorff relation \cite{Yoshida:1990}. 

For an alternative description, one can define the operator $D_L$ over the relativistic phase space spanned by $\pmb{\zeta}^T = \left(\mathbf{x}^T, \mathbf{u}^T\right)$,
\begin{displaymath}
    D_L = \left(\begin{array}{c}
        \mathbf{u}/\gamma\\
        \frac{q}{m}\left(
            \mathbf{E} + \mathbf{u}\times\mathbf{B} / \gamma
        \right)
    \end{array}\right)
    \frac{d}{d\pmb{\zeta}}.
\end{displaymath}
Then the equations of motion (\ref{eq:eqn_motion_x}) and (\ref{eq:eqn_motion_u}) can be written
\begin{displaymath}
    \frac{d\pmb{\zeta}}{dt} = D_L \pmb{\zeta},
\end{displaymath}
with the solution
\begin{displaymath}
    \pmb{\zeta}(t) = e^{t D_L} \pmb{\zeta}(0) =: \phi_{\Delta t} \pmb{\zeta}(0),
\end{displaymath}

The following operators can be used to define partial updates,
\begin{align*}
    D_x &= \frac{1}{\gamma}\left(\begin{array}{c}
        \mathbf{u}\\
        \pmb{0}
    \end{array}\right)
    \frac{d}{d\pmb{\zeta}},\\
    D_E &= \frac{q}{m}\left(\begin{array}{c}
        \pmb{0}\\
        \mathbf{E}
    \end{array}\right)
    \frac{d}{d\pmb{\zeta}},\\
    D_B &= \frac{q}{m\gamma}\left(\begin{array}{c}
        \pmb{0}\\
        \mathbf{u}\times\mathbf{B}
    \end{array}\right)
    \frac{d}{d\pmb{\zeta}}.
\end{align*}
One can easily see that these partial operators add up to the $D_L = D_x + D_E + D_B$. The GYR scheme can then be written as
\begin{equation}
\begin{split}    
    \pmb{\zeta}_{\mathrm{GYR}}(\Delta t) 
        &= e^{\frac{\Delta t}{2} D_x} e^{\frac{\Delta t}{2} D_E} 
        e^{\Delta t D_B}
        e^{\frac{\Delta t}{2} D_E} e^{\frac{\Delta t}{2} D_x} \pmb{\zeta}(0)\\
        &= e^{\Delta t D_\mathrm{GYR}}\pmb{\zeta}(0),
\end{split}\label{eq:def_d_gyr}
\end{equation}
where the last row defines the GYR operator $D_\mathrm{GYR}$. Using the Baker-Campbell-Hausdorff relations,
one can show that this operator is second-order accurate, $\Delta t D_\mathrm{GYR} = \Delta t D_L + \mathcal{O}\left(\Delta t^3\right)$. For the case of Boris, HC, and CC schemes, the exact magnetic field operator $D_B$ is replaced by the approximations
\begin{alignat*}{3}
    \mathrm{Boris}:\quad & D_B &\to \arctan\left(\frac{\beta}{\gamma_\mathrm{B}}\right)\frac{\gamma_\mathrm{B}}{\beta} D_B &= D_B + \mathcal{O}\left(\Delta t^3\right)\\
    \mathrm{HC}:\quad & D_B &\to \arctan\left(\frac{\beta}{\gamma_\mathrm{HC}}\right)\frac{\gamma_\mathrm{B}}{\beta}
    D_B &= D_B + \mathcal{O}\left(\Delta t^3\right)\\
    \mathrm{CC}:\quad & D_B &\to \arcsin\left(\frac{\beta}{\gamma_\mathrm{B}}\right)\frac{\gamma_\mathrm{B}}{\beta}
    D_B &= D_B + \mathcal{O}\left(\Delta t^3\right)
\end{alignat*}
Note that $\beta = |\pmb{\beta}|$ given in equation (\ref{eq:def_beta}) is proportional to $\Delta t$. With these replacements, equation (\ref{eq:def_d_gyr}) can be used to define the operators $D_\mathrm{Boris}$, $D_\mathrm{HC}$, and $D_\mathrm{CC}$, all of which are second-order accurate,
\begin{equation}
    \Delta t D_s = \Delta t D_L + \mathcal{O}\left(\Delta t^3\right),
\end{equation}
and volume preserving, with $s = \mathrm{Boris}, \mathrm{HC}, \mathrm{GYR}, \mathrm{CC}$. 

Thus, given these $n$-th order maps on relativistic phase space, $\phi_{\Delta t}^n = \exp(\Delta t D^n_s)$, one can formally use equations (\ref{eq:yoshida_scheme}) and (\ref{eq:yoshida_coeff}) to construct $n+2$ order schemes. This follows from the formal similarity of the equations and the fact that the order of the new scheme can be shown by using only the Baker-Campbell-Hausdorff relation. If $\phi_{\Delta t}^n$ is volume preserving in $\pmb{\zeta}$--space, then $\phi_{\Delta t}^{n+2}$ will also be volume preserving. With this, equations (\ref{eq:yoshida_scheme}) and (\ref{eq:yoshida_coeff}) provide a method to extend any of these Boris-like schemes to arbitrary order.

\subsection*{\label{sec:results_higher_order} Comparison of higher order schemes}

\begin{figure}
    \centering
    \includegraphics[width=0.49\textwidth]{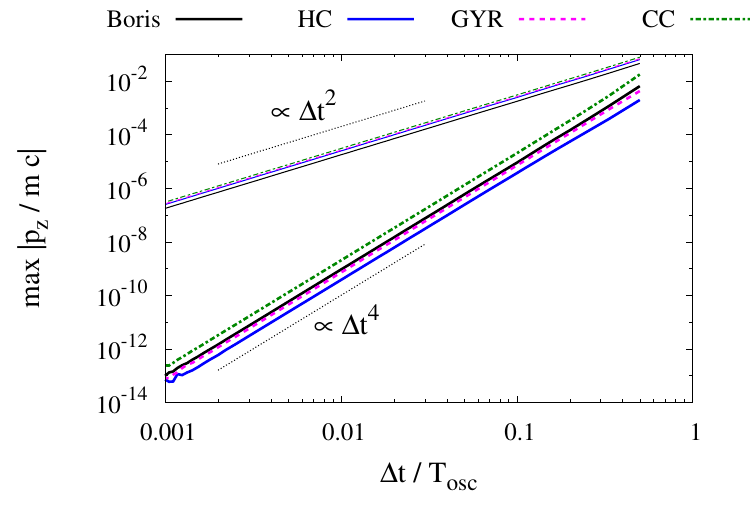}
    \includegraphics[width=0.49\textwidth]{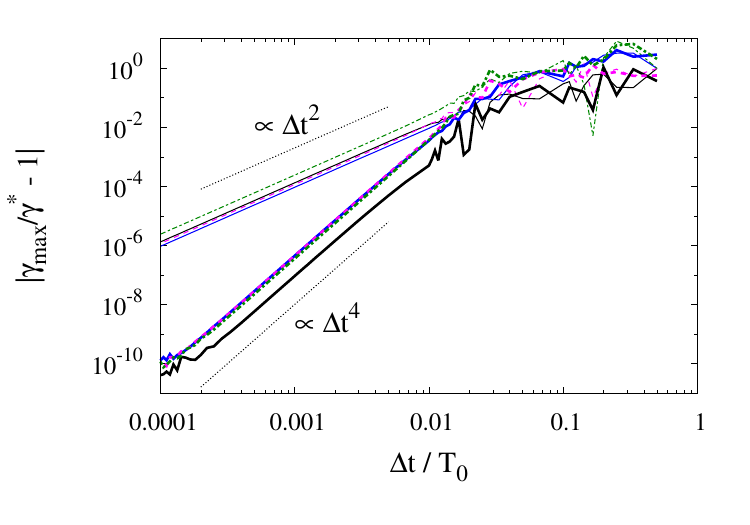}
    
    \caption{Comparison of the volume preserving type I schemes with their fourth-order variants. The left panel shows the error measure $\mathcal{E}^{\mathrm{HC}}_{p}$ of the test describe in section \ref{sec:hg_test}. The right panel shows the error measure $\mathcal{E}^{\mathrm{Wave}}_{E}$ of the test described in section \ref{sec:rel_laser}. The thin lines show the results for the original second order schemes while the thick lines show the results for the fourth order variants.
    The dotted lines indicate ideal scaling with $\Delta t^2$ and $\Delta t^4$.}
    \label{fig:scandt_y4}
\end{figure}

Two test cases were repeated for the fourth order generalisations of Boris, HC, GYR, and CC. The left panel of figure \ref{fig:scandt_y4} shows the error $\mathcal{E}^{\text{HC}}_p$ for the motion in parallel fields with spatial variation described in section \ref{sec:hg_test}. As expected, the fourth order schemes show errors reduced by at least an order of magnitude, even for the largest time steps of $\Delta t = 0.3T_c$. Fourth order convergence is clearly seen and for time steps as large as $\Delta t = 10^{-3}T_c$ all fourth order schemes approach machine precision.

The right panel of figure \ref{fig:scandt_y4} shows the error $\mathcal{E}^{\text{Wave}}_E$ for the motion in a plane wave as described in section \ref{sec:rel_laser}. Again, fourth-order convergence of the new schemes can be clearly seen for $\Delta t \le 0.01 T_0$. For larger time steps, on the other hand, where the second order schemes produce relative errors above 10\%, the fourth order schemes do not provide any overall improvement. This indicates that, while higher order schemes deliver faster convergence, they do not generally allow the use of a time step that does not resolve the underlying physics.

\section{Discussion\label{sec:discussion}}

The comparison of relativistic integration schemes in this work shows that, while some schemes tend to show smaller errors than others, there is no single integration scheme that universally outperforms the rest. IMP \cite{Ripperda:2018} is best in many cases and preserves many invariants of motion to machine precision, as is expected from an implicit scheme. However, the focus of this work is the comparison of explicit schemes, and IMP was chosen as the single implicit scheme to compare the explicit schemes against. Even then, IMP is not always the best and even shows the worst error for the magnetic moment in the magnetic bottle configuration (section \ref{sec:magnetic_bottle}). Future work is required to investigate if the otherwise good conservation properties of IMP generalise to other implicit schemes.

Among all explicit schemes, HC \cite{Higuera:2017} consistently features among the best performers. While not 
necessarily always the scheme with the lowest error, the errors are often comparable to the best explicit schemes. PL \cite{Petri:2020,Li:2021} is exact by construction for any field configuration that is constant in space and time. In principle, the time step $\Delta t$ can be chosen arbitrarily large in these situations. However, PL may suffer from the accumulation of rounding errors in some situations and the results are sensitive to the accuracy of the root finding algorithm required to solve for the particle's proper time. The other type II schemes LiLF \cite{Li:2021}, GH, GH2 \cite{Gordon:2021}, and ZZ \cite{Zhou:2024} can be viewed as variants of PL that make approximations in calculating the proper the time step $\Delta \tau$, or modify the temporal alignment of $\mathbf{x}$ and $\mathbf{u}$. These schemes also perform well in static and homogeneous field configurations. It was found that, for inhomogeneous or time dependent fields, PL, GH and GH2 reverted to first order accuracy. This is attributed to temporal alignment issues of the fields and the velocity in the calculation of the particle's proper time. Of the type II schemes, LiLF and ZZ exhibited some aspects of a second order scheme, probably due to the leap-frog time stepping of $\mathbf{x}$ and $\mathbf{u}$. For the inhomogeneous field test in case D, ZZ was one of the most accurate schemes and for the relativistic plane wave test in case F, both ZZ and LiLF performed the best among the explicit schemes. These results could, however, only be achieved when ensuring extremely low tolerances in the root finding algorithm required to solve the implicit equation for the proper time. When using a standard Newton-Raphson solver with default parameters, PL, LiLF and ZZ produced errors orders of magnitude above those of the type I schemes.

Further to the comparison of existing second order schemes, it was shown how to construct volume preserving schemes with arbitrary order from the Boris-like schemes Boris, GYR, HC, and CC. The resulting higher order schemes converge more quickly with decreasing time steps and can be useful for high accuracy calculations. However, it was also shown that no advantage is gained for large time steps, when the physical time scales are not resolved. 

No judgment was made in this work on the computational cost of the schemes. It is expected that Boris will be fastest as it contains the fewest numerical operations. HC and Vay require additional evaluations of the square root function when compared to Boris and should be only marginally slower on modern processors. All other explicit methods require the evaluation of trigonometric functions. This typically comes with computational costs. In addition, PL, LiLF and ZZ require numerical root solving to evaluate the implicit equation determining the proper time step $\Delta \tau$. While IMP does not contain any trigonometric function evaluations, a Newton iteration is performed as part of the implicit scheme. It should also be noted that, while the computational cost per time step has been a limiting factor for the speed of PIC simulations in the past, this is changing with the advent of GPU based simulation codes. On these architectures, memory bandwidth is often the limiting factor \cite{Decyk:2014}, and the so-called arithmetic intensity was found to be low. Here, an increase in the number of operations per time step could improve the arithmetic intensity without slowing down the simulation while potentially providing an improvement in accuracy.

It should be noted that all tests presented in this work assume perfect knowledge of the electromagnetic fields $\mathbf{E}$ and $\mathbf{B}$ at any time and any point in space. In realistic PIC simulations, the fields are given at discrete steps in time and space. In the case of the popular FDTD scheme the fields are given on the Yee grid \cite{Yee:1966} which is staggered in both space and time. The accuracy of particle orbits in a realistic PIC simulation will not only depend on the integration scheme itself, but also on the accuracy of the fields, positions and times at which the fields are given and the interpolation method used \cite{Tangtartharakul:2021}.

In summary, in this comparison of explicit integration schemes it was found that while the Boris scheme \cite{Boris:1970} is the default pusher in most PIC simulation codes, the scheme given by Higuera \& Cary \cite{Higuera:2017} provides a modest improvement in most test cases and a substantial improvement in the test case it was designed for, described in section \ref{sec:hg_test}. With only a small additional computational cost, this scheme stands out as a viable alternative for a general-purpose pusher. In situations where the fields are static and homogeneous the schemes given by P\'etri \cite{Petri:2020} and reformulated by Li et al \cite{Li:2021} can provide exact solutions for arbitrarily large time steps. Although not tested here, the numerical checks performed by the authors of these schemes indicate that they may also viable in more general situations in cases where the time step can be performed in proper time $\Delta \tau$ instead of $\Delta t$ in the simulation frame of reference. If more accuracy is required for typical simulation time steps, implicit methods such as the implicit midpoint method by Ripperda et al \cite{Ripperda:2018} offer an alternative at the expense of much larger computational cost. On current GPU architectures, where the computational intensity of PIC simulation is low, this might not be detrimental for performance on the whole. In addition, implicit schemes offer greater stability for larger time steps which means that the increase in computational cost may be offset by increasing the time step. It may be challenging, however, to implement a robust and efficient nonlinear solver for the implicit scheme. It was shown here that this specific implicit method performs better in most, but not all cases. Further investigation into comparing different implicit schemes is required to answer which one may be the best general purpose scheme for high accuracy simulations. Lastly, if fast convergence to the exact solution is required for small time steps, this work presents higher order versions of some Boris-like schemes which retain the conservation properties of the original schemes.

\appendix

\section*{Acknowledgements}
The author would like to thank R.~Trines for the valuable discussions and for the help in preparing the manuscript.

\section*{Declarations}
\subsection*{Competing interests}
The author has no conflicts of interest to disclose.

\subsection*{Code Availability}
The code used to generate all data of this article is freely available on GitHub at \url{https://github.com/holgerschmitz/particle-integrators}

\bibliographystyle{spphys}
\bibliography{boris_pushers}

\end{document}